\documentclass{article}

\usepackage[main, final]{neurips_2026}

\usepackage[utf8]{inputenc} 
\usepackage[T1]{fontenc}    
\usepackage{hyperref}       
\usepackage{url}            
\usepackage{booktabs}       
\usepackage{amsfonts}       
\usepackage{nicefrac}       
\usepackage{microtype}      
\usepackage{xcolor}         

\usepackage{graphicx}
\usepackage{amsmath}

\title{C3P: Contrastive promoter-protein pretraining yields representations capturing bacterial gene regulation}

\author{%
  Cameron ~Dufault\thanks{\texttt{dufaultc@cs.toronto.edu}} \\
  Department of Computer Science\\
  University of Toronto\\  
   \\
  \And
  Scott Xu \\
  Department of Computer Science \\
  University of Toronto \\
  \AND
  Alan M. Moses  \\  
  Department of Cell and Systems Biology\\
  Department of Computer Science \\
  University of Toronto \\
}

\begin{document}

\maketitle

\begin{abstract}
Despite the increasing scale of genome language models (gLMs), their ability to decode the function of regulatory sequences remains unclear. gLM pretraining relies on sequence reconstruction, which may struggle due to the noisy, rapidly evolving nature of regulatory DNA. Self-supervised contrastive approaches provide a promising alternative. Inspired by language-image architectures like CLIP, we introduce contrastive promoter-protein pretraining (C3P). By learning to align promoters to their corresponding proteins, we leverage the rich representations of proteins learned by protein language models as supervisory signal for the learning of promoter representations. After training on 88 million bacterial promoter-protein pairs, we evaluate the predictive power of C3P-learned promoter representations for inference of curated regulatory annotations, finding multi-fold improvement over leading gLMs. We also introduce zero-shot co-regulated gene retrieval, the ability to find co-regulated genes in a genome using no experimental data. We find that compared to a randomly initialized baseline, C3P training consistently provides significant zero-shot performance gains, unlike gLMs. Scaling analysis reveals the potential for further improvement as well as the efficiency of C3P, which achieved strong performance at a fraction of the training cost of leading gLMs. In addition to demonstrating that C3P training is effective for learning representations of bacterial regulatory sequences, our strong zero-shot co-regulated gene retrieval performance suggests the possibility of decoding gene regulation for millions of bacteria from their genomes alone.
\end{abstract}

\section{Introduction}

The biological effects of genes are determined by both the gene product (protein, tRNA, etc.) and when and where the product is expressed. Deep learning has revolutionized the tools for computational prediction of structure and function of proteins directly from their amino acid sequences \cite{jumper_highly_2021}, but predicting the regulation of gene expression directly from DNA sequences remains a long-standing challenge. Without regulatory context, we cannot determine from the genome sequence alone how the expression of encoded genes is orchestrated to build the cell and adapt to the environment \cite{quake_cellular_2024}. Supervised deep learning approaches have emerged for inferring regulation from DNA \cite{avsec_effective_2021,avsec_advancing_2026, linder_predicting_2025}, but require training on experimental data which is unavailable for the vast majority of organisms, in particular the millions of bacterial species \cite{louca_census-based_2019}. Despite the relative simplicity of their genomes, where the regulatory sequences are primarily contained in the regions between proteins (hereafter referred to as \textit{promoters}), for most bacterial species we know little about their gene regulation  \cite{quake_cellular_2024, baumgart_persistence_2021}.

With the vast quantity of unlabeled DNA sequencing data now available, there is clear opportunity for unsupervised approaches to help us learn regulatory function. Existing unsupervised approaches to learning from the genome have centered on language models \cite{consens_transformers_2025}. However, while protein language models (pLMs), which operate on sequences of amino acids, succeed in learning representations capturing the structure and function of proteins \cite{lin_evolutionary-scale_2023,rives_biological_2021}, multiple analyses have called into question whether existing genome language models (gLMs), which operate on DNA sequences, learn about gene regulation from their pretraining \cite{tang_evaluating_2025,vishniakov_tokenization_2025}. Demonstrations that such models learn to capture bacterial regulatory sequence function are limited (Related Work \ref{sec:relatedwork}). To our knowledge it has not been shown that gLMs learn to distinguish differentially regulated bacterial promoters.

Recent theoretical work \cite{assel_jointembedding_2025} argues that reconstruction-based tasks, such as the masked-language modeling (MLM) and next-token prediction (NTP) tasks used in training gLMs, are well-suited to the language domain, where individual tokens are information-rich and cannot be filled in based on surface level patterns. However, such tasks may be poorly suited to domains like images, where irrelevant noisy features have a large impact on pixel-level variance. Despite being discrete linear sequences of a fixed vocabulary (like natural language), regulatory DNA has long been recognized as noisy, rapidly diverging at the sequence level over evolution despite conserved function \cite{weirauch_conserved_2010, van_hijum_mechanisms_2009}. Joint-embedding tasks, which have the objective of representing alternate views of the input similarly while keeping representations of views from different inputs distant, may be better suited for learning from regulatory sequences than reconstruction-based approaches, focusing on the meaningful features distinguishing them rather than high-variance but semantically shallow features useful for reconstruction.

In this work, we introduce a self-supervised approach for learning representations of regulatory DNA sequences, contrastive promoter-protein pretraining (C3P). Inspired by language-image models like CLIP \cite{pmlr-v139-radford21a}, our approach leverages the multi-modal nature of genomic sequences, learning to align bacterial promoter representations with representations of their corresponding proteins from a pretrained pLM. This training approach is motivated by prior observations that, despite divergence in their regulatory sequences, functionally similar proteins often show similarities in their expression patterns \cite{weirauch_conserved_2010, eisen_cluster_1998}. Because pLMs capture protein function, we reasoned that alignment to their protein representations would drive similarly functioning promoters together in the representation space.

After training on 88 million promoter-protein pairs from $\sim$23,000 bacterial species, C3P models show a multi-fold improvement over leading gLMs at capturing regulatory function in bacterial promoter representations, despite having orders of magnitude fewer parameters. We evaluate this through nearest neighbour prediction of curated regulatory annotations in a held-out species, as well as performance at zero-shot co-regulated gene retrieval, an evaluation framework we introduce for exploring the potential for solving a grand challenge of regulatory genomics: inferring regulation from the genome alone. We also demonstrate that C3P scales favourably relative to gLMs, strongly improves upon randomly initialized baselines, and can be used for identifying functional sites within promoter sequences zero-shot. To summarize, our primary contributions are:
\begin{itemize}
    \item We introduce C3P, a contrastive multi-modal self-supervised approach for learning representations of regulatory sequences.
    \item We demonstrate that C3P models widely outperform leading gLMs at learning representations predictive of bacterial promoter function, and show favourable performance scaling.
    \item We introduce zero-shot co-regulated gene retrieval, an unsupervised framework for evaluating whether regulatory sequence representations capture function.
    \item We find that pretrained C3P models, unlike pretrained gLMs, consistently show significant improvement over randomly initialized baselines at zero-shot co-regulated gene retrieval.
\end{itemize}

\section{Related Work}
\label{sec:relatedwork}
\paragraph{Genome language models for bacterial regulatory genomics}

Limited demonstrations that gLMs learn bacterial regulatory function are available. Evo \cite{nguyen_sequence_2024} is a 7B parameter long-context gLM trained through NTP on a dataset containing 80,000 prokaryotic genomes at single-nucleotide resolution. On data from \textit{in vivo} assays pairing many promoters with a reporter gene \cite{lafleur_automated_2022}, the likelihood assigned by Evo to a given promoter sequence was shown to correlate with expression levels. Linear probing of Evo embeddings was also shown to be more predictive of this expression than using one-hot encoded sequences. Evo2 \cite{brixi_genome_2026} is similar but trained on $\sim$30x more data than Evo, including from eukaryotes. PromoGen2 \cite{xia_design_2026}, a gLM trained through NTP on a dataset of 1.4 million prokaryotic promoter sequences, demonstrated improved zero-shot correlation of sequence likelihood with expression over Evo. Multi-modal gLMs have also been explored, including ProDMM \cite{li_unveiling_2025} and gLM2 \cite{cornman_omg_2024}. gLM2 was trained through MLM on metagenomic sequences, tokenizing protein-coding sequences as amino acids and intergenic regions as nucleotides. Using categorical Jacobian analysis \cite{zhang_protein_2024}, gLM2 was shown to have learned the boundaries of the sigma factor binding motifs in an \textit{E. coli} promoter.

\paragraph{Self-supervised joint-embedding approaches for DNA} Like proteins \cite{Lu2020.09.04.283929} and RNA \cite{fradkin_orthrus_2026}, approaches for self-supervised contrastive learning from DNA have been shown, though very few. In DNASimCLR \cite{yang_dnasimclr_2024}, a DNA encoder was trained to learn similar representations for augmented input views, produced by random masking of nucleotides in a given DNA sequence. The most similar work to ours is RHIEPA \cite{moses_inferring_2025}, an approach for learning representations of promoters demonstrated on multiple fungal species. With RHIEPA, encoders were trained contrastively to maximize the similarity of promoters from orthologous genes, using evolution as the source of augmented views of the input \cite{lu_evolution_2020}. The trained models were shown to learn features capturing transcription factor binding motifs and produce representations that cluster by gene co-expression. Significant limitations of this approach are the need for training separate models for each species/clade to be analyzed, and for finding sets of orthologs for each gene, a difficult and computationally expensive step \cite{kilic_flexible_2020}. As it is not scalable to any given genome or gene, we do not compare C3P to RHIEPA in this work.

\paragraph{Multi-modal contrastive learning}

The objective of multi-modal contrastive learning is to connect information from different types of data, commonly images and text, by aligning positive pairs within a shared latent space. CLIP \cite{pmlr-v139-radford21a} introduced contrastive language-image pretraining by jointly training image and text encoders using a symmetric InfoNCE loss \cite{oord_representation_2019}, yielding highly transferable visual representations with strong zero-shot performance. ALIGN \cite{pmlr-v139-jia21b} showed that further scaling with noisy web data substantially improves representation quality. Whereas CLIP and ALIGN jointly optimized image and text encoders, LiT \cite{zhai2022lit} explored a more modular training strategy by freezing a pretrained vision encoder and training only the text tower to align representations. This demonstrated that strong multi-modal alignment can be achieved while preserving high-quality pretrained visual representations. Our work adapts these approaches for learning representations of regulatory DNA.

\section{C3P: contrastive promoter-protein pretraining}

\begin{figure}
    \centering
    \includegraphics[width=1.0\linewidth]{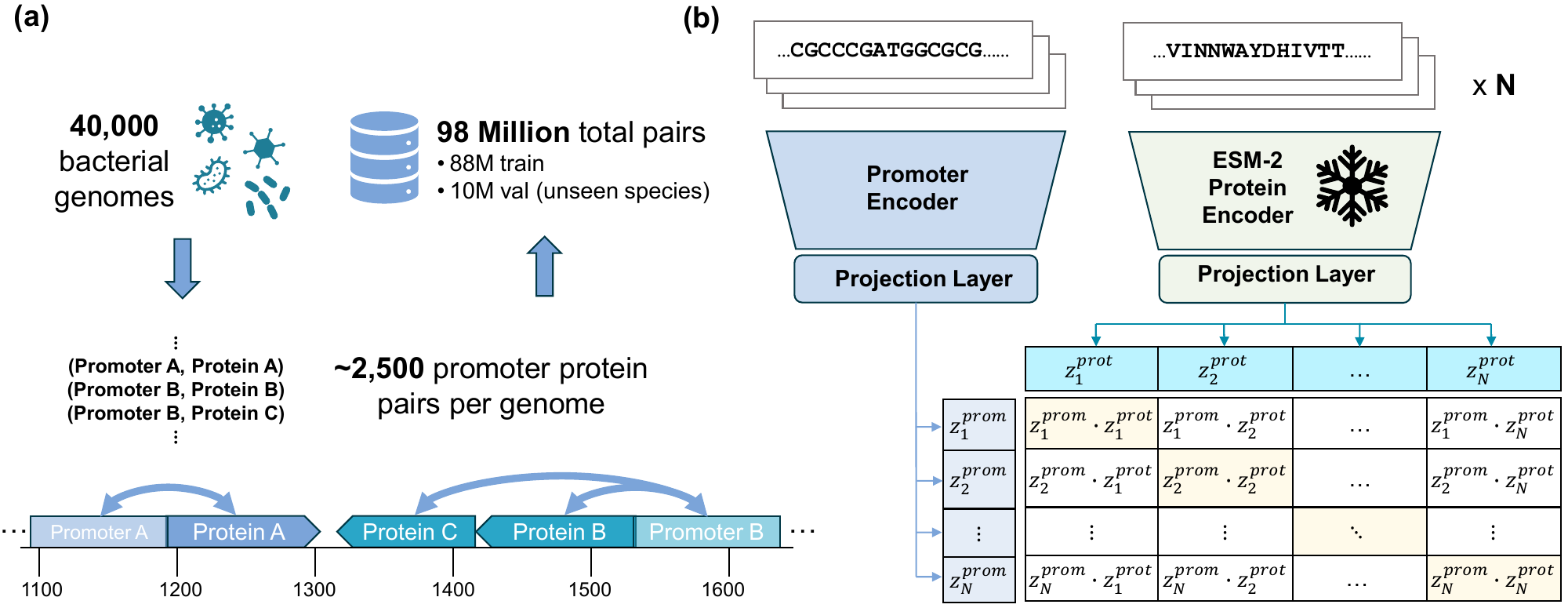}
    \caption{An overview of (a) C3P dataset creation and (b) architecture and training.}
    \label{fig:overview}
\end{figure}

\subsection{Objective and Architecture}

An overview of the C3P architecture and training task is shown in Figure \ref{fig:overview}. Inspired by language-image models such as CLIP \cite{pmlr-v139-radford21a}, the objective is to align the learned representations of promoters with the representations of their corresponding proteins.

Given a promoter-protein pair, the promoter sequence is tokenized into overlapping 3-mers and encoded by the promoter encoder, a randomly initialized transformer encoder \cite{vaswani_attention_2017}. The protein sequence is encoded by a frozen pretrained protein encoder, for which we utilize ESM2 (150M) \cite{lin_evolutionary-scale_2023}, with embeddings averaged over the sequence length. The outputs of both encoders are then passed through learnable linear projection layers to have shared dimensionality ($d_{projection} = 256$).

The model is trained with symmetric InfoNCE loss \cite{oord_representation_2019}. Given a batch of $N$ promoter-protein pairs, let $z_{i}^{prom}$ and $z_{i}^{prot}$ denote the $L_2$-normalized, projected representations of the $i$-th promoter and protein, respectively. For each batch, the model learns to maximize the cosine similarity between the $N$ correct promoter-protein pairs while minimizing the similarity between the $N^2 - N$ incorrect pairings. Where $\tau$ is a learnable temperature parameter, the per-batch loss is defined as:
\begin{equation}
    L = -\frac{1}{2N} \sum_{i=1}^{N} \left(\log \frac{\exp(z_{i}^{prom} \cdot z_{i}^{prot} / \tau)}{\sum_{j=1}^{N} \exp(z_{i}^{prom} \cdot z_{j}^{prot} / \tau)} + 
\log \frac{\exp(z_{i}^{prot} \cdot z_{i}^{prom} / \tau)}{\sum_{j=1}^{N} \exp(z_{i}^{prot} \cdot z_{j}^{prom} / \tau)}\right)
\end{equation}

\paragraph{Why align promoters to a frozen pretrained protein encoder?} 

As pretrained pLMs are known to capture functional protein features (e.g., structure \cite{lin_evolutionary-scale_2023}, localization signals \cite{sangster_zero-shot_2025}, homology \cite{rives_biological_2021}) in their representations, similarity of protein embeddings signals similar function. At the same time, proteins with similar expression patterns (implying co-regulation and therefore similarity in promoter function) are also known to often share function \cite{eisen_cluster_1998}. We reasoned that we could use protein embedding similarity as a signal that their corresponding promoters have shared function and should be represented similarly. By aligning promoter representations to pretrained protein representations, our model accomplishes this indirectly by projecting promoter sequences from similar proteins into similar regions of the representation space (see Discussion \ref{sec:discussion}). Keeping the pretrained protein encoder frozen during C3P training significantly reduces training cost, allowing for single-GPU training and for protein embeddings to be pre-computed. It also avoids the potential for representational collapse, preserving the well-structured protein latent space \cite{zhai2022lit,chen_exploring_2020}.

\subsection{Data and Model Training}

A large and diverse set of promoter-protein pairs was extracted from 40,000 bacterial genome assemblies downloaded from RefSeq \cite{goldfarb_ncbi_2025} (Appendix \ref{sec:datasampling}). To obtain positive pairs (Figure \ref{fig:overview}), for each coding gene in each genome, the non-coding region upstream of the start codon was taken as its promoter, so long as this region was at least 100 bp in length without overlapping any genes on either strand. As a rough heuristic, genes 20 bp or less in distance from another gene on the same strand were considered members of an operon, and all proteins in an operon were paired with the promoter of its first gene so long as the first gene's promoter was valid. We found this approach to accurately recover valid promoter-protein pairs from the \textit{E. coli} K-12 reference genome (94.2\% of the extracted pairs match transcription units annotated in RegulonDB \cite{salgado_regulondb_2024}, see Appendix \ref{sec:pair_validation}). 97.5 million promoter-protein pairs were obtained in total, 10.0 million of which were split into a validation dataset. To ensure generalization, validation pairs were chosen to be from species not seen during training. As prediction of \textit{E. coli} regulatory annotations is central to our evaluations, the genomes of \textit{E. coli} and closely related species were not used in creation of the training set. Appendix \ref{sec:training_data} contains a detailed description of dataset creation.

Four C3P models were trained, with 1M, 5M, 25M, and 100M trainable parameters. Each was trained for 10 epochs on the 87.5 million promoter-protein pairs in the training dataset. To reduce training costs, proteins were clustered and their embeddings pre-computed and retrieved at train time. Promoters were truncated to a maximum length of 300 bp and augmented through random cropping to a minimum of 99 bp in length. Appendix \ref{sec:modelsandtraining} contains model and training details.

\section{Experiments and Results}
  
\subsection{Multi-fold improvement over gLMs on classification of gold-standard gene regulation annotations}
\label{sec:task1}

\begin{table}[ht]
  \caption{Performance of C3P versus baseline gLMs and ESM2 (150M)  at regulon/sigmulon classification. Values represent the macro-average across all regulons with at least 5 members and all sigmulons. Uncertainty represents 2$\times$ standard error of the mean. 1-nn precision (Prec.), 1-nn Matthews correlation coefficient (MCC), and mean Average Precision@R (mAP@R)\cite{musgrave_metric_2020} shown. In brackets are trainable parameter counts.}
  
  \label{gene-representation-metrics}
  \small
  \centering
  \setlength{\tabcolsep}{3pt}
  \begin{tabular}{lcccccc}
    \toprule
    & \multicolumn{3}{c}{Regulons ($n=102$)} & \multicolumn{3}{c}{Sigmulons ($n=6$)} \\
    \cmidrule(r){2-4} \cmidrule(l){5-7}
    Model & Prec. (1-nn) & MCC (1-nn) & mAP@R  & Prec. (1-nn) & MCC (1-nn) & mAP@R \\
    \midrule
    \textbf{Ours} \\
    C3P (100M)  & \textbf{0.211$\pm.042$} & \textbf{0.184$\pm.039$} & \textbf{0.058$\pm.017$} & \textbf{0.379$\pm.116$} & \textbf{0.296$\pm.130$} & \textbf{0.077}$\pm.051$ \\
    \midrule
    \textbf{Our pLM teacher} \\
    ESM2 (150M)   & 0.088$\pm$.018 & 0.082$\pm$.020 & 0.021$\pm$.006 & 0.194$\pm$.121 & 0.114$\pm$.077 & 0.036$\pm$.047 \\
    \midrule
    \textbf{Mixed-modality gLM} \\
    gLM2 (650M)   & 0.050$\pm$.018 & 0.036$\pm$.017 & 0.009$\pm$.004& 0.144$\pm$.131 & 0.062$\pm$.101& 0.030$\pm$.040 \\    
    \midrule
    \textbf{DNA gLMs} \\
    Evo2 (7B)   & 0.037$\pm$.017 & 0.019$\pm$.011 & 0.006$\pm$.002 & 0.158$\pm$.118 & 0.073$\pm$.055 & 0.028$\pm$.038 \\
    Evo (7B)    & 0.011$\pm$.006 & -0.003$\pm$.003 & 0.003$\pm$.001& 0.075$\pm$.102 & -0.009$\pm$.014& 0.021$\pm$.035 \\
    PromoGen2 (149M) & 0.026$\pm$.016 & 0.009$\pm$.010 & 0.003$\pm$.002 & 0.102$\pm$.109 & 0.021$\pm$.020 & 0.021$\pm$.035 \\   
    \bottomrule
  \end{tabular}
\end{table}

To assess the quality of our learned promoter representations, we compared performance at predicting regulatory annotations with C3P promoter representations against those from four gLMs previously argued to learn bacterial regulatory function (Related Work \ref{sec:relatedwork}, Appendix \ref{sec:baselines}). 

We first compared their predictive power for inference of high-quality regulatory annotations from a species held out of our training data, \textit{E. coli}. RegulonDB \cite{salgado_regulondb_2024} is a curated database of \textit{E. coli} K-12 gene regulation annotations, to our knowledge the most comprehensive and up-to-date such resource for microbes. It lists \textit{regulons}, curated groups of genes experimentally validated to be regulated by the same transcription factor. It also lists \textit{sigmulons}, curated groups of genes that use the same sigma factor to initiate their transcription.

Through $k=1$ k-nearest neighbours classification, we evaluated the power of representations from each method for classifying genes by these curated groups. Promoter sequences for 2,102 \textit{E. coli} K-12 genes (standalone genes as well as the first gene of each multi-gene operon) were embedded using each method. As genes frequently belong to multiple regulons and sigmulons, we evaluated predictive performance by performing binary classification of the embeddings separately for every sigmulon ($n=6$) and for every regulon containing at least five members within our subset of genes ($n=102$).  We also measured clustering of the embedding space by each regulon and sigmulon beyond the single nearest neighbour with mean Average Precision@R (mAP@R) \cite{musgrave_metric_2020}, an information retrieval metric. Appendix \ref{sec:regulon} contains a detailed description of this evaluation.

Table \ref{gene-representation-metrics} summarizes the average performance of each method across all regulons and sigmulons. While absolute performance highlights the difficulty of this task, relative gains are substantial. C3P shows multi-fold improvement over each gLM on both regulon and sigmulon classification. Despite having orders of magnitude fewer parameters (7B vs. 100M), C3P outperforms Evo2, the best performing conventional gLM, by $\sim$9x at regulon prediction and $\sim$3x at sigmulon prediction (mAP@R performance). We also assessed whether C3P learned beyond the capabilities of its teacher by benchmarking the performance at this task using each gene's protein representations from ESM2 (150M). We again see substantial outperformance with C3P, demonstrating that despite not seeing any gene regulation data during training, C3P must have learned biological features of promoters. We additionally compared regulon prediction performance using embeddings from each method through linear probing, again finding that C3P widely outperforms all baselines (Appendix \ref{sec:linear-probing}).

\begin{figure}[h]
    \centering
    \includegraphics[width=0.90\linewidth]{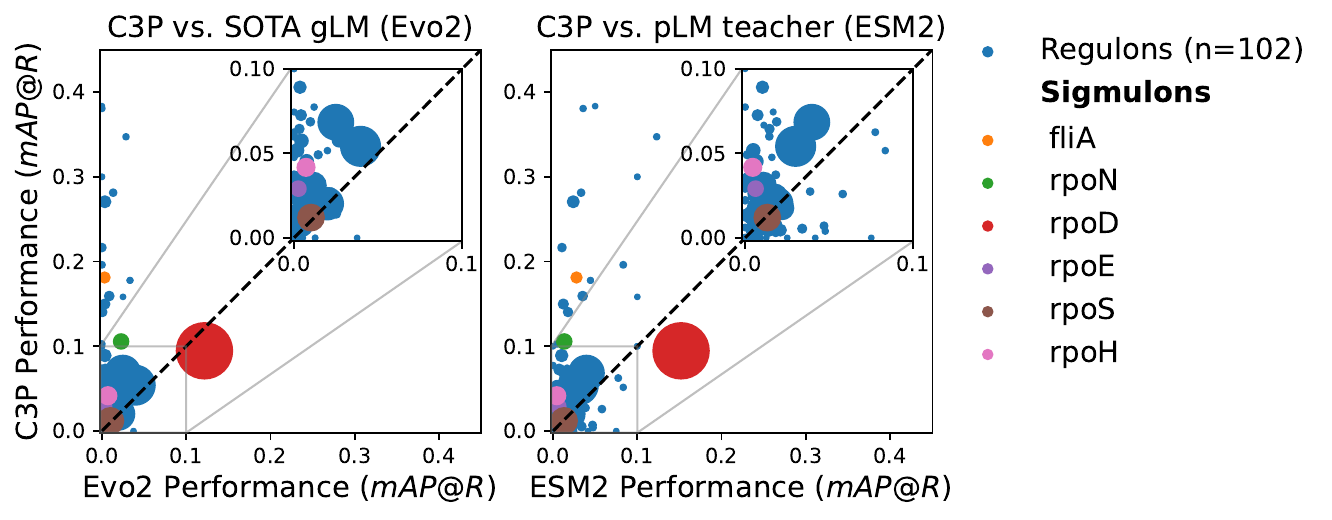}
    \caption{Comparison of C3P versus Evo2 and ESM2 (150M) per-class performance, as measured with mAP@R by the clustering of the embedding space by each \textit{E. coli} regulon ($n=102$, minimum 5 members) and sigmulon ($n=6$) in RegulonDB. Point size indicates number of genes in the class.}
    \label{fig:placeholder}
\end{figure}

We also assessed if this improvement was achieved broadly or in a handful of regulons/sigmulons. Comparing the per-class performance (by mAP@R) of C3P versus Evo2 and ESM2 (Figure \ref{fig:placeholder}), C3P outperforms or matches both baselines for nearly all regulons. For many classes this improvement is small in magnitude, but on a subset of regulons with few members C3P achieves high levels of performance. C3P also outperforms the baselines on most sigmulons, with an exception being rpoD, indicating genes regulated by the common ``housekeeping'' sigma factor $\sigma^{70}$ (see Discussion \ref{sec:discussion}).

 Finally, we repeated this evaluation in \textit{B. subtilis}, a well-studied bacterium evolutionarily distant from \textit{E. coli}. Using the curated regulatory annotations found in the DBTBS database \cite{sierro_dbtbs_2008}, we again find that C3P shows multi-fold improvement over each gLM and ESM2 (150M) (Appendix \ref{sec:dbtbs}).

\subsection{Zero-shot co-regulated gene retrieval across bacterial life with C3P promoter representations}
\label{sec:task2}

\begin{table}[ht]
  \caption{Average performance of C3P versus baseline gLMs and ESM2 (150M) at zero-shot co-regulated gene retrieval across 25 datasets. Performance on each dataset treated as a binomial proportion, uncertainty represents 2$\times$standard error of the average of the binomial proportions.}
  
  \label{zero-shot-table}
  \small
  \centering
  \setlength{\tabcolsep}{3pt}
  \begin{tabular}{lcc}
    \toprule
    Model & Top-1 Full Match Accuracy ($n=25$) & Top-1 Any Match Accuracy ($n=25$) \\
    \midrule
    \textbf{Ours} \\
    C3P (100M)  & \textbf{0.236$\pm.007$} & \textbf{0.392$\pm.008$}\\
    \midrule
    \textbf{Our pLM teacher} \\
    ESM2 (150M)   & 0.125$\pm.006$ &  0.218$\pm.007$\\
    \midrule
    \textbf{Mixed-modality gLM} \\
    gLM2 (650M)   & 0.060$\pm.005$  & 0.116$\pm.005$\\  
    \midrule
    \textbf{DNA gLMs} \\
    Evo2 (7B)   & 0.066$\pm.005$ &  0.124$\pm.006$\\
    Evo (7B)    & 0.035$\pm.003$ &  0.077$\pm.005$\\
    PromoGen2 (149M) & 0.050$\pm.004$ &  0.102$\pm.006$\\   
    \bottomrule
  \end{tabular}
\end{table}

Given our strong performance at predicting curated annotations from two evolutionarily distant species, we next evaluated whether C3P learns regulatory function across bacterial life. From the iModulonDB database \cite{rychel_imodulondb_2021} we retrieved 25 species-specific datasets of co-regulated gene groups, covering 17 unique species. Like regulons/sigmulons but inferred from gene expression experiments rather than experimentally validated, these gene groups are termed \textit{iModulons} (Appendix \ref{sec:imodulondb}).

We did not evaluate performance predicting individual iModulon membership as previously done with regulons/sigmulons (see Appendix \ref{sec:imodulondb} for justification). Instead, we use iModulons as indicators of what genes are co-regulated and introduce a novel framework for evaluating promoter representations, which we call zero-shot co-regulated gene retrieval: ability to find co-regulated genes in a genome using no experimental data. We measure the accuracy of considering a gene and its nearest neighbour in the promoter embedding space to be co-regulated through two metrics, one indicating at least a partial match in their iModulons (Top-1 Any Match) and one a complete match (Top-1 Full Match). Where $G$ is a set of $N$ genes with iModulon annotations, for every gene $g_i \in G$, where $z_i \in \mathbb{R}^{d}$ is the promoter embedding and $M_i$ is the set of iModulon annotations annotated for $g_i $, the nearest neighbour by cosine similarity is found $ g_{j^*} = \arg\max_{g_j \in G; j \neq i} \frac{z_i \cdot z_j}{\|z_i\|_2 \|z_j\|_2}$, and performance measured across all genes by:
\begin{equation}
    \text{Top-1 Any Match} = \frac{1}{N} \sum_{i=1}^{N} \mathbb{I}(M_i \cap M_{j^*} \neq \emptyset),\hspace{0.1in}\text{Top-1 Full Match} = \frac{1}{N} \sum_{i=1}^{N} \mathbb{I}(M_i = M_{j^*})
\end{equation}

\begin{figure}[h]
    \centering
    \includegraphics[width=1.0\linewidth]{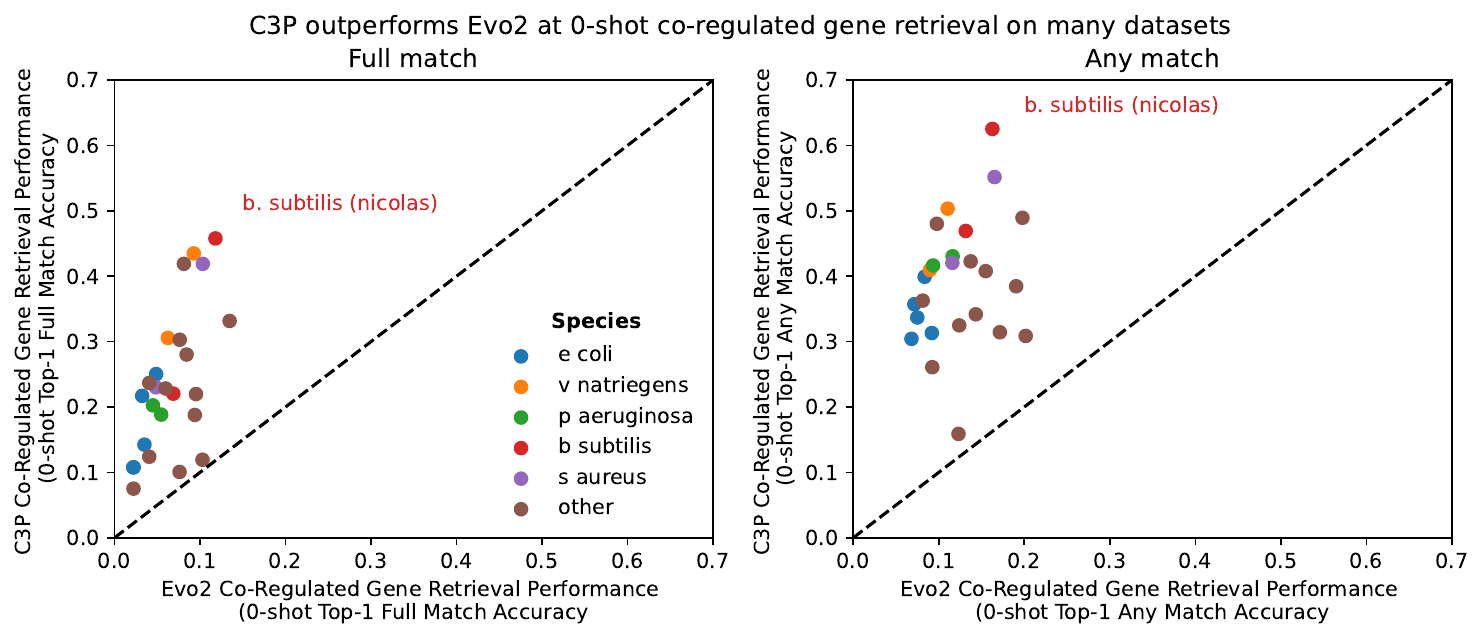}
    \caption{Comparison of C3P versus Evo2 performance measured by top-1 full match accuracy and top-1 any match accuracy on each of the 25 iModulon datasets. Datasets labeled by species.}
    \label{fig:coregbydat}
\end{figure}
For all 25 datasets we extracted promoter-protein pairs from the genome of the organism and embedded each promoter sequence with each baseline gLM as well as C3P. As in the previous task we also obtained protein embeddings from ESM2 (150M). Table \ref{zero-shot-table} gives the average performance of each method across all 25 datasets. Again, C3P widely outperforms all gLMs ($\sim$3x to 4x improvement over Evo2, the best performing gLM), as well as ESM2. We also compared the per-dataset performance of C3P with that of Evo2 to evaluate whether this improvement was seen across many species (Figure \ref{fig:coregbydat}). We see substantial improvement of C3P over Evo2 across nearly all datasets.

\begin{figure}[h]
    \centering
    \includegraphics[width=1.0\linewidth]{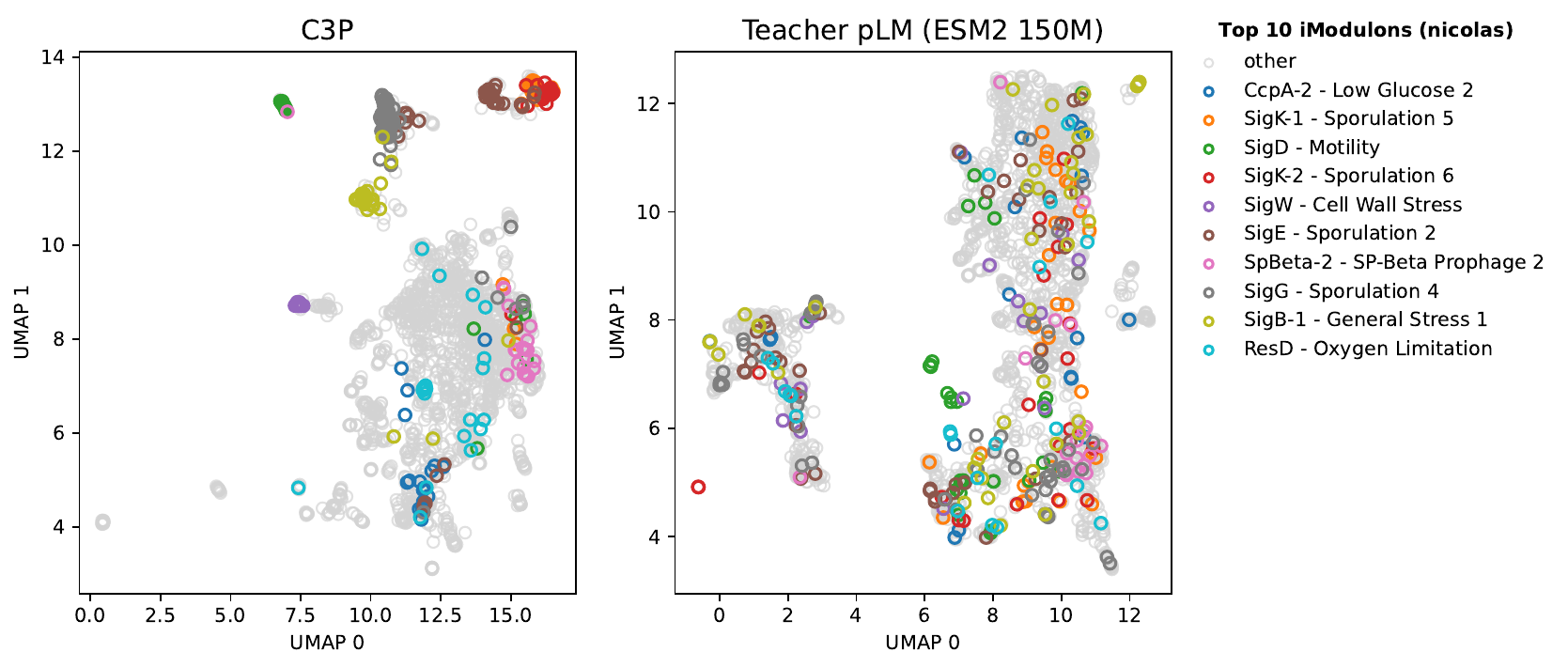}
    \caption{UMAP plots of \textit{B. subtilis} promoter embeddings from C3P and protein embeddings from ESM2 (150M). Genes labeled with one of the 10 most frequent iModulons in the \textit{nicolas} iModulonDB dataset are highlighted. Background is genes with other iModulon labels or no labels.}
    \label{fig:umap}
\end{figure}

C3P excelled at inference of \textit{B. subtilis} annotations from the \textit{nicolas} dataset, with top-1 full match and any match accuracy of 0.458 and 0.625 respectively (see Appendix \ref{sec:imodulondb} for per-dataset performance). As a qualitative evaluation, we created a UMAP \cite{mcinnes_umap_2020} plot of \textit{B. subtilis} promoter embeddings from C3P, including genes with no annotations (Figure \ref{fig:umap}). After labeling points whose sole iModulon annotations are from the 10 most common in the \textit{nicolas} dataset, a clear pattern is visible where co-regulated genes cluster together. Comparing to a UMAP plot of the same genes using protein embeddings from ESM2, our pLM teacher model, no obvious pattern of co-regulatory clustering emerges. Despite training through alignment to a frozen pLM, C3P learns features capturing regulatory function rather than protein structure.

\subsection{C3P is aligned with learning representations distinguishing differentially regulated promoters}
\label{sec:task3}

\begin{figure}[h]
    \centering
    \includegraphics[width=0.95\linewidth]{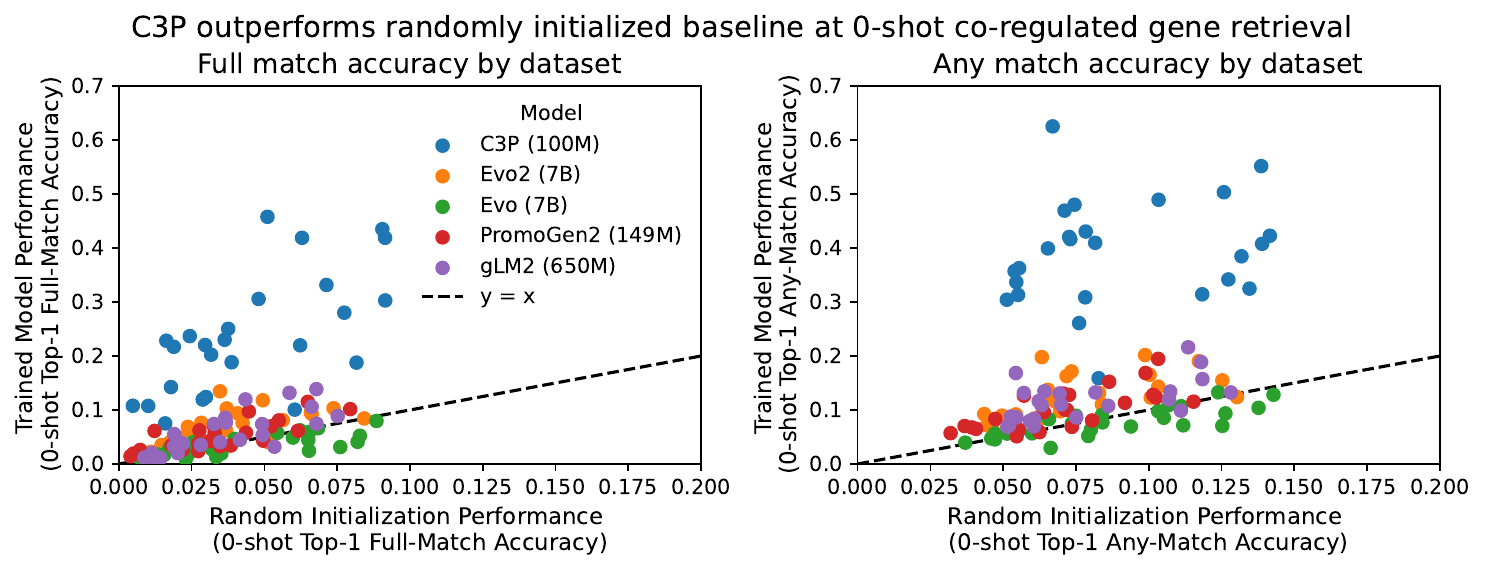}
    \caption{Comparison of the performance of C3P and each gLM baseline against their random initializations at zero-shot co-regulated gene retrieval on each of the 25 iModulonDB datasets.}
    \label{fig:random}
\end{figure}

Given the poor performance of the gLMs in our previous evaluations, we evaluated whether the pretraining of each method led to improved zero-shot co-regulated gene retrieval on each of the 25 iModulonDB datasets. We compared the performance on this task of C3P and each baseline gLM with the performance of their random initializations, which have been shown to be strong baselines for gLM performance on downstream regulatory tasks \cite{vishniakov_tokenization_2025}. On all datasets, trained C3P performance improved significantly (one-tailed two-proportion z-test $z > 1.645, p<0.05$) over a randomly initialized baseline at top-1 full and any match accuracy. While the baseline gLMs (except Evo) show evidence for improvement on many datasets (\textbf{Evo2} full: 18/25, any: 19/25; \textbf{Evo} full: 0/25, any: 0/25; \textbf{PromoGen2} full: 15/25, any: 17/25; \textbf{gLM2} full: 12/25, any: 16/25), each shows no evidence of improvement on many others. C3P demonstrates multi-fold improvement over its random baselines across nearly all datasets, while other approaches broadly show minor improvement or in some cases worsened performance (Figure \ref{fig:random}, Appendix \ref{sec:randombaseline}).

\subsection{C3P demonstrates favourable scaling for regulatory inference compared to gLMs}
\label{sec:task4}

\begin{figure}[h]
    \centering
    \includegraphics[width=0.95\linewidth]{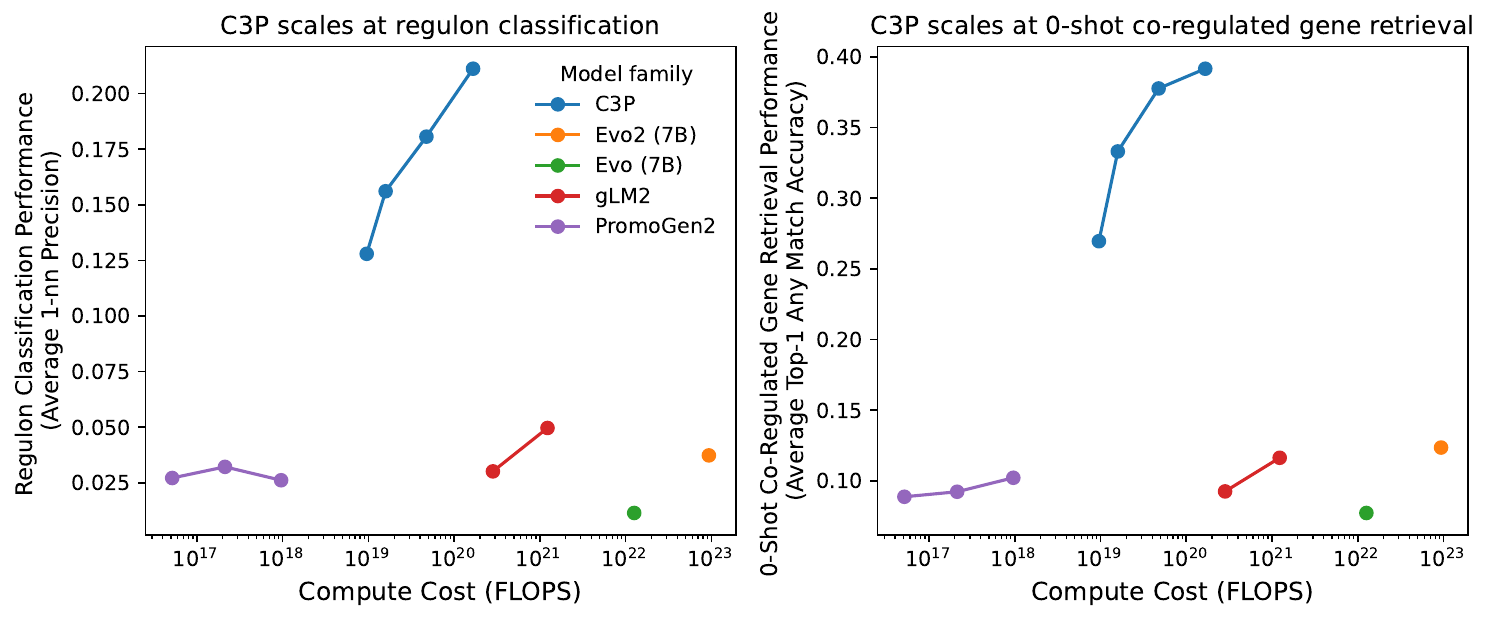}
    \caption{Scaling of regulon classification performance (average over 102 regulons) and zero-shot co-regulated gene retrieval performance (average over 25 datasets) for C3P and each gLM baseline.}
    \label{fig:flops}
\end{figure}

Noting that Evo2 showed evidence of improvement in only 3 (top-1 full match) and 2 (top-1 any match) more datasets than PromoGen2, despite having $\sim$45x more parameters, we next performed an analysis of the scaling of C3P and compared with the empirical trend of our gLM baselines. We trained four C3P models with variable numbers of learnable parameters (1M, 5M, 25M, and 100M, see Appendix \ref{sec:modelsandtraining}). We then evaluated the performance of each C3P model as well as our gLM baselines (including other models in the same family when available) and compared against their pretraining floating point operations (FLOPS, Appendix \ref{sec:scaling}). C3P models show a substantial increase in both regulon classification and zero-shot co-regulated gene retrieval performance with increasing scale (Figure \ref{fig:flops}, Appendix \ref{sec:scaling}). We also see an improved rate of scaling compared to the gLMs, where performance gains with increasing scale are minor and come at the cost of immense compute. 

\subsection{C3P enables zero-shot prediction of \textit{E. coli} promoter binding sites}
\label{sec:ismtask}

\begin{table}[ht]
  \caption{Average performance of C3P versus baseline gLMs at zero-shot prediction of binding sites across 100 \textit{E. coli} promoters. Uncertainty represents 2$\times$ standard error of the mean.}

  \label{segmentation-table}
  \small
  \centering
  \setlength{\tabcolsep}{3pt}
  \begin{tabular}{lcc}
    \toprule
    Model & Avg. AUROC & Avg. AUPRC \\
    \midrule
    \textbf{Ours} \\
    C3P (100M)  & \textbf{0.697$\pm0.039$} & \textbf{0.482$\pm0.052$}\\
    \midrule
    \textbf{Mixed-modality gLM} \\
    gLM2 (650M)   & 0.594$\pm0.031$  & 0.343$\pm0.038$\\  
    \midrule
    \textbf{DNA gLMs} \\
    Evo2 (7B)   & 0.562$\pm0.035$ &  0.325$\pm0.041$\\
    Evo (7B)    & 0.547$\pm0.035$ &  0.315$\pm0.039$\\
    PromoGen2 (149M) & 0.516$\pm0.029$ &  0.274$\pm0.039$\\   
    \bottomrule
  \end{tabular}
\end{table}

Finally, to validate that C3P learns true regulatory features, we evaluated the performance of C3P and each baseline gLM for zero-shot prediction of transcription factor and sigma factor binding sites within \textit{E. coli} promoters through \textit{in-silico} mutagenesis. For each promoter, we scored how much each model's output changes given a mutation in each position. We then evaluated the overlap of these mutation scores with the binding sites annotated in RegulonDB. For C3P, per-position mutation scores were determined by the absolute change in promoter-protein cosine similarity of the mutated promoter compared to the wild-type sequence, averaged across the three possible substitutions at each position. Average absolute change in sequence likelihood was used for the gLM baselines. Appendix  \ref{sec:segmentation} contains a detailed description of this evaluation.

Table \ref{segmentation-table} contains the results of performing this evaluation on 100 unique \textit{E. coli} promoters. C3P again widely outperforms the gLM baselines, demonstrating that the model learns to identify true functional sites in promoter sequences without training on any experimental data.

\section{Discussion}
\label{sec:discussion}

To our knowledge, it has not been shown that genome representation models learn to distinguish differentially regulated bacterial promoters. To evaluate this, we performed nearest-neighbour classification of curated regulatory annotations (Section \ref{sec:task1}), zero-shot inference of gene co-regulation in a wide array of species (Section \ref{sec:task2}), and zero-shot prediction of \textit{E. coli} promoter binding sites (Section \ref{sec:ismtask}). On these evaluations, we compared the performance of C3P against gLMs trained at a massive scale (Evo, Evo2), trained on prokaryotic promoters only (PromoGen2), and in a multi-modal fashion (gLM2). We see a multi-fold improvement over each (Table \ref{gene-representation-metrics}, Table \ref{zero-shot-table}), suggesting that scale, multi-modality, or promoter-specific training are not the key to C3P's strong performance, but rather its contrastive training regime. We believe that this is due to regulatory sequences being better suited for joint-embedding based self-supervised learning objectives than reconstruction-based objectives \cite{assel_jointembedding_2025}, as well as our use of a pretrained pLM. Despite holding out genomes from \textit{E. coli} and any closely related genera from the C3P training dataset, we see strong performance predicting \textit{E. coli} annotations in each of our evaluations. This suggests C3P performance generalizes to unseen species, although a limitation of our work is that we have not evaluated the extent of this generalization. 

C3P promoter embeddings also widely outperformed ESM2 (150M) protein embeddings, revealing that C3P does not simply learn a lossy copy of protein embeddings, but instead learns new features relevant to promoter function. This is further evidenced by the qualitative differences in clustering visible in a UMAP plot of \textit{B. subtilis} embeddings from both methods (Figure \ref{fig:umap}), and the strong performance of C3P at zero-shot prediction of \textit{E. coli} promoter binding sites (Table \ref{segmentation-table}). A limitation of this work is that we have not performed interpretability analysis of the model's learned features. Given our strong results and the current inability of experts to determine regulatory function from sequence alone, doing so may reveal currently unknown rules of regulatory logic.

The biological inspiration of C3P was that functionally related proteins often share expression patterns, and thus because pretrained pLMs capture protein function we could learn to drive representations of similarly functioning promoters together based on the similarity of their corresponding protein embeddings. There are many ways to make use of this insight we did not explore, such as unfreezing the ESM2 model and learning a joint promoter-protein latent space. Another more direct approach than C3P may be to use an approach like that of LGSimCLR \cite{elbanani2022languageguided}, where positive promoter pairs for contrastive training are created based on protein embedding similarity. However, such an approach would reduce the rich features captured in protein representations \cite{sangster_zero-shot_2025} to similarity scores.

Results from our scaling analysis indicate that C3P scales favourably compared to gLMs (Section \ref{sec:task4}). A caveat to this result is that C3P utilizes a pretrained protein encoder, while the gLMs we compared against were trained from scratch. Our scaling analysis was also limited to C3P models with varying trainable parameter sizes. We note that further increasing the size of the promoter encoder may improve training loss, but risks memorization of promoters rather than learning regulatory features. With millions of bacterial genomes in GenBank \cite{sayers_genbank_2025} there is instead opportunity for data scaling. We have not evaluated how sensitive our results are to the specific protein encoder used. Models like ESM-C \cite{esm2024cambrian}, larger and trained on more proteins than ESM2 (150M), may improve performance. 

While we have limited the scope of this work to bacterial promoters, clear avenues of future work are extending our approach to eukaryotic genomes and to other regulatory regions (e.g., enhancers). For single-celled eukaryotes like yeast, our approach may be readily adaptable. Alternative approaches will be needed for higher-order organisms, like humans, where gene regulation is not largely determined by short regions upstream of genes. To accurately capture gene regulation, such approaches must model long-range regulatory interactions, cell-type specific effects, and alternative splicing.

Demonstrations that gLMs learn the function of bacterial regulatory sequences have thus far largely been based on prediction of expression data derived from experimental assays where many distinct promoter sequences are paired with a reporter gene, capturing promoter strength \cite{nguyen_sequence_2024,xia_design_2026}. While the purpose of this work was not to learn representations predictive of promoter strength, we also evaluated our performance at this task following the approach of Evo \cite{nguyen_sequence_2024}, where linear models are trained to predict synthetic promoter activity. Interestingly, we find the 1M and 5M parameter C3P models show strongest performance on this evaluation, outperforming all gLM baselines and matching the performance of a state-of-the-art supervised biophysical model \cite{lafleur_automated_2022}, while the larger C3P models (25M and 100M) underperformed. When performing this evaluation on a dataset where wild-type promoter strength is measured \cite{johns_metagenomic_2018}, C3P does not consistently outperform the gLM baselines, but we again see decreased performance with increased C3P model size. Appendix \ref{sec:evotask} details this evaluation.

Because C3P is trained to distinguish promoters from one another, it may fail to learn common features less useful for that task. An example of this may be the poor performance seen at clustering of promoters by the very common rpoD sigmulon (Figure \ref{fig:placeholder}). A significant drawback of C3P compared to autoregressive gLMs is the inability to generate new sequences, a focus of existing approaches \cite{nguyen_sequence_2024,xia_design_2026}. A multi-task learning setup \cite{fradkin_orthrus_2026} may enable this, but we have not explored this possibility.

\section{Conclusion}

Genome language models are known to struggle to decode the function of regulatory sequences \cite{tang_evaluating_2025,vishniakov_tokenization_2025}. Inspired by the success of multi-modal contrastive alignment in the language-image domain \cite{pmlr-v139-radford21a}, we introduced C3P, a novel self-supervised approach for learning representations of regulatory sequences. Using C3P, we trained a promoter encoder on 88 million bacterial promoter-protein pairs and evaluated its performance on nearest-neighbour prediction of gold-standard regulatory annotations, zero-shot co-regulated gene retrieval across 25 diverse bacterial datasets, and zero-shot prediction of \textit{E. coli} promoter binding sites. We demonstrated that C3P representations are substantially more predictive of gene regulation than leading gLMs, despite requiring orders of magnitude less compute. Furthermore, we confirmed that C3P consistently leads to significant performance improvement over a randomly initialized baseline, unlike gLMs. The success of C3P at zero-shot co-regulated gene retrieval suggests the possibility of inferring the regulatory networks of entirely unstudied species, an important step towards the ability to build complete models of cells entirely from the genome \cite{quake_cellular_2024}.

\newpage
\begin{ack}
We thank David Knowles, Alex Lu, Philip Fradkin, Andrew Duncan, Ami Sangster, Rain Jin, and Micaela Consens for feedback on the manuscript. SX is employed by Waabi; this work was carried out independently of that employment. AMM acknowledges support from NSERC and the Canada Research Chairs program. Computing resources used in preparing this research were provided, in part, by the Province of Ontario, the Government of Canada through CIFAR, and companies sponsoring the Vector Institute.
\end{ack}

\bibliographystyle{plain}
\bibliography{neurips_2026}

@article{lin_evolutionary-scale_2023,
	title = {Evolutionary-scale prediction of atomic-level protein structure with a language model},
	volume = {379},
	url = {https://www.science.org/doi/10.1126/science.ade2574},
	doi = {10.1126/science.ade2574},
	number = {6637},
	urldate = {2026-04-09},
	journal = {Science},
	publisher = {American Association for the Advancement of Science},
	author = {Lin, Zeming and Akin, Halil and Rao, Roshan and Hie, Brian and Zhu, Zhongkai and Lu, Wenting and Smetanin, Nikita and Verkuil, Robert and Kabeli, Ori and Shmueli, Yaniv and dos Santos Costa, Allan and Fazel-Zarandi, Maryam and Sercu, Tom and Candido, Salvatore and Rives, Alexander},
	month = mar,
	year = {2023},
	pages = {1123--1130},
}

@article{brixi_genome_2026,
	title = {Genome modelling and design across all domains of life with {Evo} 2},
	copyright = {2026 The Author(s)},
	issn = {1476-4687},
	url = {https://www.nature.com/articles/s41586-026-10176-5},
	doi = {10.1038/s41586-026-10176-5},
	language = {en},
	urldate = {2026-04-09},
	journal = {Nature},
	publisher = {Nature Publishing Group},
	author = {Brixi, Garyk and Durrant, Matthew G. and Ku, Jerome and Naghipourfar, Mohsen and Poli, Michael and Sun, Gwanggyu and Brockman, Greg and Chang, Daniel and Fanton, Alison and Gonzalez, Gabriel A. and King, Samuel H. and Li, David B. and Merchant, Aditi T. and Nguyen, Eric and Ricci-Tam, Chiara and Romero, David W. and Schmok, Jonathan C. and Taghibakhshi, Ali and Vorontsov, Anton and Yang, Brandon and Deng, Myra and Gorton, Liv and Nguyen, Nam and Wang, Nicholas K. and Pearce, Michael T. and Simon, Elana and Adams, Etowah and Amador, Zachary J. and Ashley, Euan A. and Baccus, Stephen A. and Dai, Haoyu and Dillmann, Steven and Ermon, Stefano and Guo, Daniel and Herschl, Michael H. and Ilango, Rajesh and Janik, Ken and Lu, Amy X. and Mehta, Reshma and Mofrad, Mohammad R. K. and Ng, Madelena Y. and Pannu, Jaspreet and Ré, Christopher and St. John, John and Sullivan, Jeremy and Tey, Joseph and Viggiano, Ben and Zhu, Kevin and Zynda, Greg and Balsam, Daniel and Collison, Patrick and Costa, Anthony B. and Hernandez-Boussard, Tina and Ho, Eric and Liu, Ming-Yu and McGrath, Thomas and Powell, Kimberly and Pinglay, Sudarshan and Burke, Dave P. and Goodarzi, Hani and Hsu, Patrick D. and Hie, Brian L.},
	month = mar,
	year = {2026},
	pages = {1--13},
}

@article{xia_design_2026,
	title = {Design prokaryotic cis-regulatory elements using language model},
	volume = {54},
	issn = {1362-4962},
	url = {https://doi.org/10.1093/nar/gkag122},
	doi = {10.1093/nar/gkag122},
	number = {4},
	urldate = {2026-04-08},
	journal = {Nucleic Acids Research},
	author = {Xia, Yan and Sun, Jinyuan and Du, Xiaowen and Liang, Zeyu and Wu, Xin and Shi, Wenyu and Shao, Bin and Guo, Shuyuan and Huo, Yi-Xin},
	month = feb,
	year = {2026},
	pages = {gkag122},
}

@article{weirauch_conserved_2010,
	title = {Conserved expression without conserved regulatory sequence: the more things change, the more they stay the same},
	volume = {26},
	issn = {0168-9525},
	shorttitle = {Conserved expression without conserved regulatory sequence},
	url = {https://www.cell.com/trends/genetics/abstract/S0168-9525(09)00250-9},
	doi = {10.1016/j.tig.2009.12.002},
	language = {English},
	number = {2},
	urldate = {2026-04-08},
	journal = {Trends in Genetics},
	publisher = {Elsevier},
	author = {Weirauch, Matthew T. and Hughes, Timothy R.},
	month = feb,
	year = {2010},
	pages = {66--74},
}

@article{lafleur_automated_2022,
	title = {Automated model-predictive design of synthetic promoters to control transcriptional profiles in bacteria},
	volume = {13},
	copyright = {2022 The Author(s)},
	issn = {2041-1723},
	url = {https://www.nature.com/articles/s41467-022-32829-5},
	doi = {10.1038/s41467-022-32829-5},
	language = {en},
	number = {1},
	urldate = {2026-04-08},
	journal = {Nature Communications},
	publisher = {Nature Publishing Group},
	author = {LaFleur, Travis L. and Hossain, Ayaan and Salis, Howard M.},
	month = sep,
	year = {2022},
	pages = {5159},
}

@article{oleary_exploring_2024,
	title = {Exploring and retrieving sequence and metadata for species across the tree of life with {NCBI} {Datasets}},
	volume = {11},
	copyright = {2024 This is a U.S. Government work and not under copyright protection in the US; foreign copyright protection may apply},
	issn = {2052-4463},
	url = {https://www.nature.com/articles/s41597-024-03571-y},
	doi = {10.1038/s41597-024-03571-y},
	language = {en},
	number = {1},
	urldate = {2026-04-08},
	journal = {Scientific Data},
	publisher = {Nature Publishing Group},
	author = {O’Leary, Nuala A. and Cox, Eric and Holmes, J. Bradley and Anderson, W. Ray and Falk, Robert and Hem, Vichet and Tsuchiya, Mirian T. N. and Schuler, Gregory D. and Zhang, Xuan and Torcivia, John and Ketter, Anne and Breen, Laurie and Cothran, Jonathan and Bajwa, Hena and Tinne, Jovany and Meric, Peter A. and Hlavina, Wratko and Schneider, Valerie A.},
	month = jul,
	year = {2024},
	pages = {732},
}

@article{goldfarb_ncbi_2025,
	title = {{NCBI} {RefSeq}: reference sequence standards through 25 years of curation and annotation},
	volume = {53},
	issn = {1362-4962},
	shorttitle = {{NCBI} {RefSeq}},
	url = {https://doi.org/10.1093/nar/gkae1038},
	doi = {10.1093/nar/gkae1038},
	number = {D1},
	urldate = {2026-04-08},
	journal = {Nucleic Acids Research},
	author = {Goldfarb, Tamara and Kodali, Vamsi K and Pujar, Shashikant and Brover, Vyacheslav and Robbertse, Barbara and Farrell, Catherine M and Oh, Dong-Ha and Astashyn, Alexander and Ermolaeva, Olga and Haddad, Diana and Hlavina, Wratko and Hoffman, Jinna and Jackson, John D and Joardar, Vinita S and Kristensen, David and Masterson, Patrick and McGarvey, Kelly M and McVeigh, Richard and Mozes, Eyal and Murphy, Michael R and Schafer, Susan S and Souvorov, Alexander and Spurrier, Brett and Strope, Pooja K and Sun, Hanzhen and Vatsan, Anjana R and Wallin, Craig and Webb, David and Brister, J Rodney and Hatcher, Eneida and Kimchi, Avi and Klimke, William and Marchler-Bauer, Aron and Pruitt, Kim D and Thibaud-Nissen, Françoise and Murphy, Terence D},
	month = jan,
	year = {2025},
	pages = {D243--D257},
}

@article{tang_evaluating_2025,
	title = {Evaluating the representational power of pre-trained {DNA} language models for regulatory genomics},
	volume = {26},
	issn = {1474-760X},
	url = {https://doi.org/10.1186/s13059-025-03674-8},
	doi = {10.1186/s13059-025-03674-8},
	language = {en},
	number = {1},
	urldate = {2026-04-08},
	journal = {Genome Biology},
	author = {Tang, Ziqi and Somia, Nirali and Yu, Yiyang and Koo, Peter K.},
	month = jul,
	year = {2025},
	pages = {203},
}

@article{jumper_highly_2021,
	title = {Highly accurate protein structure prediction with {AlphaFold}},
	volume = {596},
	copyright = {2021 The Author(s)},
	issn = {1476-4687},
	url = {https://www.nature.com/articles/s41586-021-03819-2},
	doi = {10.1038/s41586-021-03819-2},
	language = {en},
	number = {7873},
	urldate = {2026-04-08},
	journal = {Nature},
	publisher = {Nature Publishing Group},
	author = {Jumper, John and Evans, Richard and Pritzel, Alexander and Green, Tim and Figurnov, Michael and Ronneberger, Olaf and Tunyasuvunakool, Kathryn and Bates, Russ and Žídek, Augustin and Potapenko, Anna and Bridgland, Alex and Meyer, Clemens and Kohl, Simon A. A. and Ballard, Andrew J. and Cowie, Andrew and Romera-Paredes, Bernardino and Nikolov, Stanislav and Jain, Rishub and Adler, Jonas and Back, Trevor and Petersen, Stig and Reiman, David and Clancy, Ellen and Zielinski, Michal and Steinegger, Martin and Pacholska, Michalina and Berghammer, Tamas and Bodenstein, Sebastian and Silver, David and Vinyals, Oriol and Senior, Andrew W. and Kavukcuoglu, Koray and Kohli, Pushmeet and Hassabis, Demis},
	month = aug,
	year = {2021},
	pages = {583--589},
}

@article{quake_cellular_2024,
	title = {The cellular dogma},
	volume = {187},
	issn = {0092-8674, 1097-4172},
	url = {https://www.cell.com/cell/abstract/S0092-8674(24)01211-X},
	doi = {10.1016/j.cell.2024.10.029},
	language = {English},
	number = {23},
	urldate = {2026-04-07},
	journal = {Cell},
	publisher = {Elsevier},
	author = {Quake, Stephen R.},
	month = nov,
	year = {2024},
	pages = {6421--6423},
}

@article{nguyen_sequence_2024,
	title = {Sequence modeling and design from molecular to genome scale with {Evo}},
	volume = {386},
	issn = {0036-8075, 1095-9203},
	url = {https://www.science.org/doi/10.1126/science.ado9336},
	doi = {10.1126/science.ado9336},
	language = {en},
	number = {6723},
	urldate = {2025-01-01},
	journal = {Science},
	author = {Nguyen, Eric and Poli, Michael and Durrant, Matthew G. and Kang, Brian and Katrekar, Dhruva and Li, David B. and Bartie, Liam J. and Thomas, Armin W. and King, Samuel H. and Brixi, Garyk and Sullivan, Jeremy and Ng, Madelena Y. and Lewis, Ashley and Lou, Aaron and Ermon, Stefano and Baccus, Stephen A. and Hernandez-Boussard, Tina and Ré, Christopher and Hsu, Patrick D. and Hie, Brian L.},
	month = nov,
	year = {2024},
	pages = {eado9336},
}

@article{louca_census-based_2019,
	title = {A census-based estimate of {Earth}'s bacterial and archaeal diversity},
	volume = {17},
	issn = {1545-7885},
	url = {https://dx.plos.org/10.1371/journal.pbio.3000106},
	doi = {10.1371/journal.pbio.3000106},
	language = {en},
	number = {2},
	urldate = {2025-01-02},
	journal = {PLOS Biology},
	author = {Louca, Stilianos and Mazel, Florent and Doebeli, Michael and Parfrey, Laura Wegener},
	editor = {Jansson, Janet K.},
	month = feb,
	year = {2019},
	pages = {e3000106},
}

@article{steinegger_mmseqs2_2017,
	title = {{MMseqs2} enables sensitive protein sequence searching for the analysis of massive data sets},
	volume = {35},
	copyright = {2017 Springer Nature America, Inc.},
	issn = {1546-1696},
	url = {https://www.nature.com/articles/nbt.3988},
	doi = {10.1038/nbt.3988},
	language = {en},
	number = {11},
	urldate = {2025-11-19},
	journal = {Nature Biotechnology},
	publisher = {Nature Publishing Group},
	author = {Steinegger, Martin and Söding, Johannes},
	month = nov,
	year = {2017},
	pages = {1026--1028},
}

@article{salgado_regulondb_2024,
	title = {{RegulonDB} v12.0: a comprehensive resource of transcriptional regulation in {E}. coli {K}-12},
	volume = {52},
	issn = {0305-1048},
	shorttitle = {{RegulonDB} v12.0},
	url = {https://doi.org/10.1093/nar/gkad1072},
	doi = {10.1093/nar/gkad1072},
	number = {D1},
	urldate = {2025-11-30},
	journal = {Nucleic Acids Research},
	author = {Salgado, Heladia and Gama-Castro, Socorro and Lara, Paloma and Mejia-Almonte, Citlalli and Alarcón-Carranza, Gabriel and López-Almazo, Andrés G and Betancourt-Figueroa, Felipe and Peña-Loredo, Pablo and Alquicira-Hernández, Shirley and Ledezma-Tejeida, Daniela and Arizmendi-Zagal, Lizeth and Mendez-Hernandez, Francisco and Diaz-Gomez, Ana K and Ochoa-Praxedis, Elizabeth and Muñiz-Rascado, Luis J and García-Sotelo, Jair S and Flores-Gallegos, Fanny A and Gómez, Laura and Bonavides-Martínez, César and del Moral-Chávez, Víctor M and Hernández-Alvarez, Alfredo J and Santos-Zavaleta, Alberto and Capella-Gutierrez, Salvador and Gelpi, Josep Lluis and Collado-Vides, Julio},
	month = jan,
	year = {2024},
	pages = {D255--D264},
}

@article{consens_transformers_2025,
	title = {Transformers and genome language models},
	volume = {7},
	copyright = {2025 Springer Nature Limited},
	issn = {2522-5839},
	url = {https://www.nature.com/articles/s42256-025-01007-9},
	doi = {10.1038/s42256-025-01007-9},
	language = {en},
	number = {3},
	urldate = {2025-12-01},
	journal = {Nature Machine Intelligence},
	publisher = {Nature Publishing Group},
	author = {Consens, Micaela E. and Dufault, Cameron and Wainberg, Michael and Forster, Duncan and Karimzadeh, Mehran and Goodarzi, Hani and Theis, Fabian J. and Moses, Alan and Wang, Bo},
	month = mar,
	year = {2025},
	pages = {346--362},
}

@article{moses_inferring_2025,
	title = {Inferring fungal cis-regulatory networks from genome sequences via unsupervised and interpretable representation learning},
	issn = {1943-2631},
	doi = {10.1093/genetics/iyaf209},
	language = {eng},
	journal = {Genetics},
	author = {Moses, Alan M. and Stajich, Jason E. and Gasch, Audrey P. and Knowles, David A.},
	month = sep,
	year = {2025},
	pages = {iyaf209},
}

@misc{mcinnes_umap_2020,
	title = {{UMAP}: {Uniform} {Manifold} {Approximation} and {Projection} for {Dimension} {Reduction}},
	shorttitle = {{UMAP}},
	url = {http://arxiv.org/abs/1802.03426},
	doi = {10.48550/arXiv.1802.03426},
	urldate = {2025-12-02},
	publisher = {arXiv},
	author = {McInnes, Leland and Healy, John and Melville, James},
	month = sep,
	year = {2020},
	note = {arXiv:1802.03426 [stat]},
}

@inproceedings{vaswani_attention_2017,
	title = {Attention is {All} you {Need}},
	volume = {30},
	url = {https://proceedings.neurips.cc/paper_files/paper/2017/hash/3f5ee243547dee91fbd053c1c4a845aa-Abstract.html},
	urldate = {2026-01-19},
	booktitle = {Advances in {Neural} {Information} {Processing} {Systems}},
	publisher = {Curran Associates, Inc.},
	author = {Vaswani, Ashish and Shazeer, Noam and Parmar, Niki and Uszkoreit, Jakob and Jones, Llion and Gomez, Aidan N and Kaiser, Łukasz and Polosukhin, Illia},
	year = {2017},
}

@inproceedings{
    vishniakov_tokenization_2025,
    title={Tokenization to Transfer: Do Genomic Foundation Models Learn Good Representations?},
    author={Kirill Vishniakov and Karthik Viswanathan and Aleksandr Medvedev and Praveenkumar Kanithi and Marco AF Pimentel and Ronnie Rajan and Shadab Khan},
    booktitle={The Fourteenth International Conference on Learning Representations},
    year={2026},
    url={https://openreview.net/forum?id=4UY1NHG5Ge}
}

@inproceedings{assel_jointembedding_2025,
	title = {Joint‑{Embedding} vs {Reconstruction}: {Provable} {Benefits} of {Latent} {Space} {Prediction} for {Self}‑{Supervised} {Learning}},
	shorttitle = {Joint‑{Embedding} vs {Reconstruction}},
	url = {https://openreview.net/forum?id=UOaLsgn5wb},
	language = {en},
	urldate = {2026-05-01},
	booktitle = {The {Thirty}-ninth {Annual} {Conference} on {Neural} {Information} {Processing} {Systems}},
	author = {Assel, Hugues Van and Ibrahim, Mark and Biancalani, Tommaso and Regev, Aviv and Balestriero, Randall},
	month = oct,
	year = {2025},
}

@inproceedings{
cornman_omg_2024,
title={The {OMG} dataset: An Open MetaGenomic corpus for mixed-modality genomic language modeling},
author={Andre Cornman and Jacob West-Roberts and Antonio Pedro Camargo and Simon Roux and Martin Beracochea and Milot Mirdita and Sergey Ovchinnikov and Yunha Hwang},
booktitle={The Thirteenth International Conference on Learning Representations},
year={2025},
url={https://openreview.net/forum?id=jlzNb1iWs3}
}

@misc{oord_representation_2019,
	title = {Representation {Learning} with {Contrastive} {Predictive} {Coding}},
	url = {http://arxiv.org/abs/1807.03748},
	doi = {10.48550/arXiv.1807.03748},
	urldate = {2026-05-01},
	publisher = {arXiv},
	author = {Oord, Aaron van den and Li, Yazhe and Vinyals, Oriol},
	month = jan,
	year = {2019},
	note = {arXiv:1807.03748 [cs]},
}

@article{rychel_imodulondb_2021,
	title = {{iModulonDB}: a knowledgebase of microbial transcriptional regulation derived from machine learning},
	volume = {49},
	issn = {0305-1048},
	shorttitle = {{iModulonDB}},
	url = {https://doi.org/10.1093/nar/gkaa810},
	doi = {10.1093/nar/gkaa810},
	number = {D1},
	urldate = {2026-05-01},
	journal = {Nucleic Acids Research},
	author = {Rychel, Kevin and Decker, Katherine and Sastry, Anand V and Phaneuf, Patrick V and Poudel, Saugat and Palsson, Bernhard O},
	month = jan,
	year = {2021},
	pages = {D112--D120},
}

@article{yang_dnasimclr_2024,
	title = {{DNASimCLR}: a contrastive learning-based deep learning approach for gene sequence data classification},
	volume = {25},
	issn = {1471-2105},
	shorttitle = {{DNASimCLR}},
	url = {https://doi.org/10.1186/s12859-024-05955-8},
	doi = {10.1186/s12859-024-05955-8},
	language = {en},
	number = {1},
	urldate = {2026-05-01},
	journal = {BMC Bioinformatics},
	author = {Yang, Minghao and Wang, Zehua and Yan, Zizhuo and Wang, Wenxiang and Zhu, Qian and Jin, Changlong},
	month = oct,
	year = {2024},
	pages = {328},
}

@inproceedings{musgrave_metric_2020,
	address = {Cham},
	title = {A {Metric} {Learning} {Reality} {Check}},
	isbn = {978-3-030-58595-2},
	doi = {10.1007/978-3-030-58595-2_41},
	language = {en},
	booktitle = {Computer {Vision} – {ECCV} 2020},
	publisher = {Springer International Publishing},
	author = {Musgrave, Kevin and Belongie, Serge and Lim, Ser-Nam},
	year = {2020},
	pages = {681--699},
}

@inproceedings{pmlr-v139-radford21a,
  title = 	 {Learning Transferable Visual Models From Natural Language Supervision},
  author =       {Radford, Alec and Kim, Jong Wook and Hallacy, Chris and Ramesh, Aditya and Goh, Gabriel and Agarwal, Sandhini and Sastry, Girish and Askell, Amanda and Mishkin, Pamela and Clark, Jack and Krueger, Gretchen and Sutskever, Ilya},
  booktitle = 	 {Proceedings of the 38th International Conference on Machine Learning},
  pages = 	 {8748--8763},
  year = 	 {2021},
  volume = 	 {139},
  series = 	 {Proceedings of Machine Learning Research},
  month = 	 {18--24 Jul},
  publisher =    {PMLR},
  url = 	 {https://proceedings.mlr.press/v139/radford21a.html}
}

@InProceedings{pmlr-v139-jia21b,
  title = 	 {Scaling Up Visual and Vision-Language Representation Learning With Noisy Text Supervision},
  author =       {Jia, Chao and Yang, Yinfei and Xia, Ye and Chen, Yi-Ting and Parekh, Zarana and Pham, Hieu and Le, Quoc and Sung, Yun-Hsuan and Li, Zhen and Duerig, Tom},
  booktitle = 	 {Proceedings of the 38th International Conference on Machine Learning},
  pages = 	 {4904--4916},
  year = 	 {2021},
  volume = 	 {139},
  series = 	 {Proceedings of Machine Learning Research},
  month = 	 {18--24 Jul},
  publisher =    {PMLR},
  url = 	 {https://proceedings.mlr.press/v139/jia21b.html}
}

@InProceedings{zhai2022lit,
  author={Zhai, Xiaohua and Wang, Xiao and Mustafa, Basil and Steiner, Andreas and Keysers, Daniel and Kolesnikov, Alexander and Beyer, Lucas},
  booktitle={2022 IEEE/CVF Conference on Computer Vision and Pattern Recognition (CVPR)}, 
  title={LiT: Zero-Shot Transfer with Locked-image text Tuning}, 
  year={2022},
  volume={},
  number={},
  pages={18102-18112},
  doi={10.1109/CVPR52688.2022.01759}
}

@article{avsec_advancing_2026,
	title = {Advancing regulatory variant effect prediction with {AlphaGenome}},
	volume = {649},
	copyright = {2026 The Author(s)},
	issn = {1476-4687},
	url = {https://www.nature.com/articles/s41586-025-10014-0},
	doi = {10.1038/s41586-025-10014-0},
	language = {en},
	number = {8099},
	urldate = {2026-05-01},
	journal = {Nature},
	publisher = {Nature Publishing Group},
	author = {Avsec, Žiga and Latysheva, Natasha and Cheng, Jun and Novati, Guido and Taylor, Kyle R. and Ward, Tom and Bycroft, Clare and Nicolaisen, Lauren and Arvaniti, Eirini and Pan, Joshua and Thomas, Raina and Dutordoir, Vincent and Perino, Matteo and De, Soham and Karollus, Alexander and Gayoso, Adam and Sargeant, Toby and Mottram, Anne and Wong, Lai Hong and Drotár, Pavol and Kosiorek, Adam and Senior, Andrew and Tanburn, Richard and Applebaum, Taylor and Basu, Souradeep and Hassabis, Demis and Kohli, Pushmeet},
	month = jan,
	year = {2026},
	pages = {1206--1218},
}

@article{avsec_effective_2021,
	title = {Effective gene expression prediction from sequence by integrating long-range interactions},
	volume = {18},
	copyright = {2021 The Author(s)},
	issn = {1548-7105},
	url = {https://www.nature.com/articles/s41592-021-01252-x},
	doi = {10.1038/s41592-021-01252-x},
	language = {en},
	number = {10},
	urldate = {2026-05-01},
	journal = {Nature Methods},
	publisher = {Nature Publishing Group},
	author = {Avsec, Žiga and Agarwal, Vikram and Visentin, Daniel and Ledsam, Joseph R. and Grabska-Barwinska, Agnieszka and Taylor, Kyle R. and Assael, Yannis and Jumper, John and Kohli, Pushmeet and Kelley, David R.},
	month = oct,
	year = {2021},
	pages = {1196--1203},
}

@article{sangster_zero-shot_2025,
	title = {Zero-shot segmentation using embeddings from a protein language model identifies functional regions in the human proteome},
	volume = {21},
	issn = {1553-7358},
	url = {https://journals.plos.org/ploscompbiol/article?id=10.1371/journal.pcbi.1012929},
	doi = {10.1371/journal.pcbi.1012929},
	language = {en},
	number = {11},
	urldate = {2026-05-01},
	journal = {PLOS Computational Biology},
	publisher = {Public Library of Science},
	author = {Sangster, Ami G. and Dufault, Cameron and Qu, Haoning and Le, Denise and Forman-Kay, Julie D. and Moses, Alan M.},
	month = nov,
	year = {2025},
	pages = {e1012929},
}

@article{kilic_flexible_2020,
	title = {Flexible comparative genomics of prokaryotic transcriptional regulatory networks},
	volume = {21},
	issn = {1471-2164},
	url = {https://doi.org/10.1186/s12864-020-06838-x},
	doi = {10.1186/s12864-020-06838-x},
	language = {en},
	number = {5},
	urldate = {2026-05-01},
	journal = {BMC Genomics},
	author = {Kılıç, Sefa and Sánchez-Osuna, Miquel and Collado-Padilla, Antonio and Barbé, Jordi and Erill, Ivan},
	month = dec,
	year = {2020},
	pages = {466},
}

@article{rives_biological_2021,
	title = {Biological structure and function emerge from scaling unsupervised learning to 250 million protein sequences},
	volume = {118},
	url = {https://www.pnas.org/doi/full/10.1073/pnas.2016239118},
	doi = {10.1073/pnas.2016239118},
	number = {15},
	urldate = {2026-05-01},
	journal = {Proceedings of the National Academy of Sciences},
	publisher = {Proceedings of the National Academy of Sciences},
	author = {Rives, Alexander and Meier, Joshua and Sercu, Tom and Goyal, Siddharth and Lin, Zeming and Liu, Jason and Guo, Demi and Ott, Myle and Zitnick, C. Lawrence and Ma, Jerry and Fergus, Rob},
	month = apr,
	year = {2021},
	pages = {e2016239118},
}

@article{zhang_protein_2024,
	title = {Protein language models learn evolutionary statistics of interacting sequence motifs},
	volume = {121},
	url = {https://www.pnas.org/doi/10.1073/pnas.2406285121},
	doi = {10.1073/pnas.2406285121},
	number = {45},
	urldate = {2026-05-01},
	journal = {Proceedings of the National Academy of Sciences},
	publisher = {Proceedings of the National Academy of Sciences},
	author = {Zhang, Zhidian and Wayment-Steele, Hannah K. and Brixi, Garyk and Wang, Haobo and Kern, Dorothee and Ovchinnikov, Sergey},
	month = nov,
	year = {2024},
	pages = {e2406285121},
}

@misc{lu_evolution_2020,
	title = {Evolution {Is} {All} {You} {Need}: {Phylogenetic} {Augmentation} for {Contrastive} {Learning}},
	shorttitle = {Evolution {Is} {All} {You} {Need}},
	url = {http://arxiv.org/abs/2012.13475},
	doi = {10.48550/arXiv.2012.13475},
	urldate = {2026-05-01},
	publisher = {arXiv},
	author = {Lu, Amy X. and Lu, Alex X. and Moses, Alan},
	month = dec,
	year = {2020},
	note = {arXiv:2012.13475 [q-bio]},
}

@inproceedings{elbanani2022languageguided,
  title={{Learning Visual Representations via Language-Guided Sampling}},
  author={El Banani, Mohamed and Desai, Karan and Johnson, Justin},
  booktitle={CVPR},
  year={2023},
}

@article{sastry_escherichia_2019,
	title = {The {Escherichia} coli transcriptome mostly consists of independently regulated modules},
	volume = {10},
	copyright = {2019 The Author(s)},
	issn = {2041-1723},
	url = {https://www.nature.com/articles/s41467-019-13483-w},
	doi = {10.1038/s41467-019-13483-w},
	language = {en},
	number = {1},
	urldate = {2026-05-03},
	journal = {Nature Communications},
	publisher = {Nature Publishing Group},
	author = {Sastry, Anand V. and Gao, Ye and Szubin, Richard and Hefner, Ying and Xu, Sibei and Kim, Donghyuk and Choudhary, Kumari Sonal and Yang, Laurence and King, Zachary A. and Palsson, Bernhard O.},
	month = dec,
	year = {2019},
	pages = {5536},
}

@article{esm2024cambrian,
  title={ESM Cambrian: Revealing the mysteries of proteins with unsupervised learning},
  author={ESM Team and others},
  journal={EvolutionaryScale Website},
  year={2024}
}

@article{sayers_genbank_2025,
	title = {{GenBank} 2025 update},
	volume = {53},
	issn = {1362-4962},
	url = {https://doi.org/10.1093/nar/gkae1114},
	doi = {10.1093/nar/gkae1114},
	number = {D1},
	urldate = {2026-05-03},
	journal = {Nucleic Acids Research},
	author = {Sayers, Eric W and Cavanaugh, Mark and Frisse, Linda and Pruitt, Kim D and Schneider, Valerie A and Underwood, Beverly A and Yankie, Linda and Karsch-Mizrachi, Ilene},
	month = jan,
	year = {2025},
	pages = {D56--D61},
}

@article{stripedhyena,
  title={Stripedhyena: Moving beyond transformers with hybrid signal processing models},
  author={Poli, Michael and Wang, Jue and Massaroli, Stefano and Quesnelle, Jeffrey and Carlow, Ryan and Nguyen, Eric and Thomas, Armin},
  journal={GitHub repository},
  volume={12},
  year={2023},
}

@misc{kaplan2020scalinglawsneurallanguage,
      title={Scaling Laws for Neural Language Models}, 
      author={Jared Kaplan and Sam McCandlish and Tom Henighan and Tom B. Brown and Benjamin Chess and Rewon Child and Scott Gray and Alec Radford and Jeffrey Wu and Dario Amodei},
      year={2020},
      eprint={2001.08361},
      archivePrefix={arXiv},
      primaryClass={cs.LG},
      url={https://arxiv.org/abs/2001.08361}, 
      note={arXiv:2001.08361 [cs]},
}

@misc{su2021roformerenhancedtransformerrotary,
      title={RoFormer: Enhanced Transformer with Rotary Position Embedding}, 
      author={Jianlin Su and Yu Lu and Shengfeng Pan and Bo Wen and Yunfeng Liu},
      year={2021},
      eprint={2104.09864v1},
      archivePrefix={arXiv},
      primaryClass={cs.CL},
      url={https://arxiv.org/abs/2104.09864v1}, 
      note={arXiv:2104.09864v1 [cs]},
}

@inproceedings{
loshchilov2019adamw,
title={Decoupled Weight Decay Regularization},
author={Ilya Loshchilov and Frank Hutter},
booktitle={International Conference on Learning Representations},
year={2019},
url={https://openreview.net/forum?id=Bkg6RiCqY7},
}

@article{yu_multiplexed_2021,
	title = {Multiplexed characterization of rationally designed promoter architectures deconstructs combinatorial logic for {IPTG}-inducible systems},
	volume = {12},
	copyright = {2021 The Author(s)},
	issn = {2041-1723},
	url = {https://www.nature.com/articles/s41467-020-20094-3},
	doi = {10.1038/s41467-020-20094-3},
	language = {en},
	number = {1},
	urldate = {2026-05-06},
	journal = {Nature Communications},
	publisher = {Nature Publishing Group},
	author = {Yu, Timothy C. and Liu, Winnie L. and Brinck, Marcia S. and Davis, Jessica E. and Shek, Jeremy and Bower, Grace and Einav, Tal and Insigne, Kimberly D. and Phillips, Rob and Kosuri, Sriram and Urtecho, Guillaume},
	month = jan,
	year = {2021},
	pages = {325},
}

@article{hossain_automated_2020,
	title = {Automated design of thousands of nonrepetitive parts for engineering stable genetic systems},
	volume = {38},
	copyright = {2020 The Author(s), under exclusive licence to Springer Nature America, Inc.},
	issn = {1546-1696},
	url = {https://www.nature.com/articles/s41587-020-0584-2},
	doi = {10.1038/s41587-020-0584-2},
	language = {en},
	number = {12},
	urldate = {2026-05-06},
	journal = {Nature Biotechnology},
	publisher = {Nature Publishing Group},
	author = {Hossain, Ayaan and Lopez, Eriberto and Halper, Sean M. and Cetnar, Daniel P. and Reis, Alexander C. and Strickland, Devin and Klavins, Eric and Salis, Howard M.},
	month = dec,
	year = {2020},
	pages = {1466--1475},
}

@article{urtecho_systematic_2019,
	title = {Systematic {Dissection} of {Sequence} {Elements} {Controlling} {$\sigma$}70 {Promoters} {Using} a {Genomically} {Encoded} {Multiplexed} {Reporter} {Assay} in {Escherichia} coli},
	volume = {58},
	issn = {0006-2960},
	url = {https://doi.org/10.1021/acs.biochem.7b01069},
	doi = {10.1021/acs.biochem.7b01069},
	number = {11},
	urldate = {2026-05-06},
	journal = {Biochemistry},
	publisher = {American Chemical Society},
	author = {Urtecho, Guillaume and Tripp, Arielle D. and Insigne, Kimberly D. and Kim, Hwangbeom and Kosuri, Sriram},
	month = mar,
	year = {2019},
	pages = {1539--1551},
}

@article{li_unveiling_2025,
  title={Harnessing A Unified Multi-modal Sequence Modeling to unveil Protein-DNA Interdependency},
  author={Li, Mingchen and Ren, Yuchen and Ye, Peng and Cheng, Jiabei and Ma, Xinzhu and Cai, Yuchen and Ouyang, Wanli and Zhong, Bozitao and Wu, Banghao and Dong, Nanqing and others},
  journal={bioRxiv},
  pages={2025--02},
  year={2025},
  publisher={Cold Spring Harbor Laboratory}
}

@inproceedings{ramzi_robust_2021,
	title = {Robust and {Decomposable} {Average} {Precision} for {Image} {Retrieval}},
	volume = {34},
	url = {https://proceedings.neurips.cc/paper_files/paper/2021/hash/c622c085c04eadc473f08541b255320e-Abstract.html},
	urldate = {2026-05-06},
	booktitle = {Advances in {Neural} {Information} {Processing} {Systems}},
	publisher = {Curran Associates, Inc.},
	author = {Ramzi, Elias and Thome, Nicolas and Rambour, Clément and Audebert, Nicolas and Bitot, Xavier},
	year = {2021},
	pages = {23569--23581},
}

@article{scikit-learn,
  title={Scikit-learn: Machine Learning in {P}ython},
  author={Pedregosa, F. and Varoquaux, G. and Gramfort, A. and Michel, V.
          and Thirion, B. and Grisel, O. and Blondel, M. and Prettenhofer, P.
          and Weiss, R. and Dubourg, V. and Vanderplas, J. and Passos, A. and
          Cournapeau, D. and Brucher, M. and Perrot, M. and Duchesnay, E.},
  journal={Journal of Machine Learning Research},
  volume={12},
  pages={2825--2830},
  year={2011}
}

@article {Lu2020.09.04.283929,
	author = {Lu, Amy X. and Zhang, Haoran and Ghassemi, Marzyeh and Moses, Alan},
	title = {Self-Supervised Contrastive Learning of Protein Representations By Mutual Information Maximization},
	elocation-id = {2020.09.04.283929},
	year = {2020},
	doi = {10.1101/2020.09.04.283929},
	publisher = {Cold Spring Harbor Laboratory},
	URL = {https://www.biorxiv.org/content/early/2020/11/10/2020.09.04.283929},
	eprint = {https://www.biorxiv.org/content/early/2020/11/10/2020.09.04.283929.full.pdf},
	journal = {bioRxiv}
}

@article{fradkin_orthrus_2026,
	title = {Orthrus: toward evolutionary and functional {RNA} foundation models},
	copyright = {2026 The Author(s), under exclusive licence to Springer Nature America, Inc.},
	issn = {1548-7105},
	shorttitle = {Orthrus},
	url = {https://www.nature.com/articles/s41592-026-03064-3},
	doi = {10.1038/s41592-026-03064-3},
	language = {en},
	urldate = {2026-05-06},
	journal = {Nature Methods},
	publisher = {Nature Publishing Group},
	author = {Fradkin, Philip and Shi, Ruian “Ian” and Dalal, Taykhoom and Isaev, Keren and Frey, Brendan J. and Lee, Leo J. and Morris, Quaid and Wang, Bo},
	month = apr,
	year = {2026},
	pages = {1--11},
}

@article{baumgart_persistence_2021,
	title = {Persistence and plasticity in bacterial gene regulation},
	volume = {18},
	copyright = {2021 The Author(s), under exclusive licence to Springer Nature America, Inc.},
	issn = {1548-7105},
	url = {https://www.nature.com/articles/s41592-021-01312-2},
	doi = {10.1038/s41592-021-01312-2},
	language = {en},
	number = {12},
	urldate = {2026-05-07},
	journal = {Nature Methods},
	publisher = {Nature Publishing Group},
	author = {Baumgart, Leo A. and Lee, Ji Eun and Salamov, Asaf and Dilworth, David J. and Na, Hyunsoo and Mingay, Matthew and Blow, Matthew J. and Zhang, Yu and Yoshinaga, Yuko and Daum, Chris G. and O’Malley, Ronan C.},
	month = dec,
	year = {2021},
	pages = {1499--1505},
}

@article{eisen_cluster_1998,
	title = {Cluster analysis and display of genome-wide expression patterns},
	volume = {95},
	url = {https://www.pnas.org/doi/full/10.1073/pnas.95.25.14863},
	doi = {10.1073/pnas.95.25.14863},
	number = {25},
	urldate = {2026-05-07},
	journal = {Proceedings of the National Academy of Sciences},
	publisher = {Proceedings of the National Academy of Sciences},
	author = {Eisen, Michael B. and Spellman, Paul T. and Brown, Patrick O. and Botstein, David},
	month = dec,
	year = {1998},
	pages = {14863--14868},
}

@misc{chen_exploring_2020,
	title = {Exploring {Simple} {Siamese} {Representation} {Learning}},
	url = {http://arxiv.org/abs/2011.10566},
	doi = {10.48550/arXiv.2011.10566},
	urldate = {2026-05-07},
	publisher = {arXiv},
	author = {Chen, Xinlei and He, Kaiming},
	month = nov,
	year = {2020},
	note = {arXiv:2011.10566 [cs]},
}

@article{linder_predicting_2025,
	title = {Predicting {RNA}-seq coverage from {DNA} sequence as a unifying model of gene regulation},
	volume = {57},
	issn = {1061-4036, 1546-1718},
	url = {https://www.nature.com/articles/s41588-024-02053-6},
	doi = {10.1038/s41588-024-02053-6},
	language = {en},
	number = {4},
	urldate = {2026-05-22},
	journal = {Nature Genetics},
	author = {Linder, Johannes and Srivastava, Divyanshi and Yuan, Han and Agarwal, Vikram and Kelley, David R.},
	month = apr,
	year = {2025},
	pages = {949--961},
}

@article{alnajar_phylogenomics_2017,
	title = {Phylogenomics and comparative genomic studies delineate six main clades within the family {Enterobacteriaceae} and support the reclassification of several polyphyletic members of the family},
	volume = {54},
	issn = {15671348},
	url = {https://linkinghub.elsevier.com/retrieve/pii/S1567134817302186},
	doi = {10.1016/j.meegid.2017.06.024},
	language = {en},
	urldate = {2026-07-28},
	journal = {Infection, Genetics and Evolution},
	author = {Alnajar, Seema and Gupta, Radhey S.},
	month = oct,
	year = {2017},
	pages = {108--127},
}

@article{sierro_dbtbs_2008,
	title = {{DBTBS}: a database of transcriptional regulation in {Bacillus} subtilis containing upstream intergenic conservation information},
	volume = {36},
	copyright = {http://creativecommons.org/licenses/by-nc/2.0/uk/},
	issn = {0305-1048, 1362-4962},
	shorttitle = {{DBTBS}},
	url = {https://academic.oup.com/nar/article-lookup/doi/10.1093/nar/gkm910},
	doi = {10.1093/nar/gkm910},
	language = {en},
	number = {suppl\_1},
	urldate = {2026-07-28},
	journal = {Nucleic Acids Research},
	author = {Sierro, Nicolas and Makita, Yuko and De Hoon, Michiel and Nakai, Kenta},
	month = jan,
	year = {2008},
	pages = {D93--D96},
}

@article{hou_understanding_2026,
	title = {Understanding language model scaling for protein fitness prediction},
	volume = {6},
	issn = {2662-8457},
	url = {https://www.nature.com/articles/s43588-026-01010-z},
	doi = {10.1038/s43588-026-01010-z},
	language = {en},
	number = {7},
	urldate = {2026-10-19},
	journal = {Nature Computational Science},
	author = {Hou, Chao and Liu, Di and Zafar, Aziz and Shen, Yufeng},
	month = jul,
	year = {2026},
	pages = {778--788},
}

@article{johns_metagenomic_2018,
	title = {Metagenomic mining of regulatory elements enables programmable species-selective gene expression},
	volume = {15},
	issn = {1548-7091, 1548-7105},
	url = {https://www.nature.com/articles/nmeth.4633},
	doi = {10.1038/nmeth.4633},
	language = {en},
	number = {5},
	urldate = {2026-10-27},
	journal = {Nature Methods},
	author = {Johns, Nathan I and Gomes, Antonio L C and Yim, Sung Sun and Yang, Anthony and Blazejewski, Tomasz and Smillie, Christopher S and Smith, Mark B and Alm, Eric J and Kosuri, Sriram and Wang, Harris H},
	month = may,
	year = {2018},
	pages = {323--329},
}

@article{van_hijum_mechanisms_2009,
	title = {Mechanisms and {Evolution} of {Control} {Logic} in {Prokaryotic} {Transcriptional} {Regulation}},
	volume = {73},
	issn = {1092-2172, 1098-5557},
	url = {https://journals.asm.org/doi/10.1128/MMBR.00037-08},
	doi = {10.1128/MMBR.00037-08},
	language = {en},
	number = {3},
	urldate = {2026-11-05},
	journal = {Microbiology and Molecular Biology Reviews},
	author = {Van Hijum, Sacha A. F. T. and Medema, Marnix H. and Kuipers, Oscar P.},
	month = sep,
	year = {2009},
	pages = {481--509},
}


\newpage

\appendix

\section{Data and Code Availability}

All data used in this work was sourced from publicly available databases. Codes for downloading data, extracting promoter-protein pairs, and training C3P models are available at \href{https://github.com/dufaultc/contrastive-promoter-protein-pretraining}{https://github.com/dufaultc/contrastive-promoter-protein-pretraining}. Trained models are available at \href{https://huggingface.co/dufaultc/contrastive-promoter-protein-pretraining}{https://huggingface.co/dufaultc/contrastive-promoter-protein-pretraining}.

\section{Dataset creation}
\subsection{Genome sampling and diversity}
\label{sec:datasampling}

\begin{table}[h]
\centering
\caption{Summary of taxonomic diversity among the 35,878 bacterial genome assemblies in the train dataset. Median number of members of the same taxa, as well as max number of members of the same taxa is shown.}
\label{tab:bacterial_diversity}
\begin{tabular}{lccc}
\toprule
\textbf{Taxonomic rank} & \textbf{Number of unique taxa} & \textbf{Median \# of members} & \textbf{Max \# of members} \\ \midrule
Phylum & 60 & 22.5  & 14,470\\
Class & 133 & 12 & 6,283\\
Order & 318 & 13& 2,302\\
Family & 820 &9&508 \\
Genus & 3,546 & 3 &169\\
Species & 22,756 & 1 &7\\ 
\bottomrule
\end{tabular}
\end{table}

Using NCBI Datasets \cite{oleary_exploring_2024}, metadata for all 472,415 (excluding atypical) bacterial genomes available in the RefSeq database \cite{goldfarb_ncbi_2025} (\url{https://www.ncbi.nlm.nih.gov/refseq/}, free for public use) was downloaded on February 18, 2026. A diverse subset of these genomes was then selected in the following manner: 80,000 genomes were selected with maximal family-level diversity (most families represented with the least genomes sharing a family). From these, 60,000 were selected with maximal genus-level diversity, and finally 40,000 with maximal species-level diversity were selected and their assemblies downloaded (genome sequence, annotation features, and protein files). This approach was taken to maximize taxonomic diversity at multiple levels. The 40,000 genomes were then split into train and validation datasets at the species level. Of the unique species represented in the 40,000, 10\% were randomly selected (with \textit{E. coli} being specifically added) and genomes belonging to this 10\% of species were used to build the validation set (4,072 genomes). Finally, the 50 remaining genomes belonging to genera comprising the "Escherichia clade" \cite{alnajar_phylogenomics_2017} (Escherichia, Shigella, Citrobacter, and Salmonella) were removed from the training set. This was done in order to ensure our evaluations on \textit{E. coli} annotations reflect performance on an unseen clade. Table \ref{tab:bacterial_diversity} gives an overview of the diversity of the 35,878 genomes in the train dataset.

\subsection{Promoter-protein pair extraction}
\label{sec:training_data}
For each genome in the train and validation dataset, as well as those used in our evaluations, promoter-protein pairs were extracted according to the following process: 
\begin{itemize}
    \item Annotation features (GFF) file parsed sequentially, with the locations (beginning and end) of annotated genes (all types) saved.
    \item GFF again parsed sequentially. For each coding sequence (CDS) encountered, the distance between the start codon of the CDS and the nearest annotated gene is found, referencing the previously saved locations. Based on that distance we do the following if the CDS is on the positive strand:
    \begin{itemize}
        \item If that distance is at least 100 bp in length, the non-coding region covering that distance (up to a max of 512 bp) is recorded as the promoter for that CDS. A positive strand operon is started and the promoter is associated with it.
        \item If that distance is 20 bp or less, and greater than -20 bp, and the previous annotation encountered started a positive strand operon or is a member of a positive strand operon, the promoter which was associated with the first member of the operon is recorded as the promoter for this CDS.
        \item Otherwise, we move to the next CDS, and end any operon if one has begun.
    \end{itemize}
    \item If the CDS is on the negative strand, we find the distance between the end of the CDS and the nearest nearby annotated gene, and based on that distance do the following:
    \begin{itemize}
        \item If that distance is 20 bp or less, and greater than -20 bp, a negative strand operon is started if one has not yet been started, and the CDS added as an operon member.    
        \item If that distance is at least 100 bp in length, the non-coding region covering that distance (up to a max of 512 bp) is reverse complemented and recorded as the promoter for that CDS. If a negative strand operon is active, all CDS in that operon are also associated with this promoter.
        \item Otherwise, we move to the next CDS, and discard any entries added to the active operon if one exists.
    \end{itemize}    
\end{itemize}

Each promoter and associated protein are then added to the dataset as a pair, where each pair represents one gene. The identifier of the gene is also recorded for each pair, along with a flag indicating whether a gene was added because it was part of an operon but was not the first member of that operon.

\subsection{Validation of promoter-protein pair extraction heuristic}
\label{sec:pair_validation}
To validate that the above approach extracts valid promoter-protein pairs, we compare the pairs we extract from the \textit{E. coli} K-12 reference genome with the transcription units (groups of genes transcribed from the same promoter) annotated in RegulonDB. All transcription unit annotations were downloaded on July 24 2026.  

Of the 2,815 (out of the 2,902 total) extracted pairs belonging to genes with any transcription unit annotated, we find that 94.2\% can be matched to transcription units in RegulonDB. This accuracy is 95.9\% for the 2,045 pairs we predict to be standalone/first-operon member genes, and 89.6\% for the 770 we predict to be secondary operon members (meaning they share the promoter of a first-operon member).

\section{C3P Models and Training}
\label{sec:modelsandtraining}

\begin{table}[h]
  \caption{Variable architecture features of C3P models and compute required for training. As training was split into multiple runs, times are approximate (representative of one round of training with the listed GPUs used), and do not account for time required for pre-computation of protein embeddings.}
  \label{architecture-table}
  \centering
  \setlength{\tabcolsep}{3pt}
  \begin{tabular}{lcccccc}
    \toprule
    Trainable parameters & Heads & Layers & $d_{promoter}$ & Train time & GPU & Training FLOPS \\
    \midrule
    1M & 4 & 4 & 128 & 1 day & L40s & $9.57 \times 10^{18}$ \\
    5M & 8 & 6 & 256 & 3 days  & L40s& $1.59 \times 10^{19}$ \\
    25M & 8 & 8 & 512 &  2.5 days & H100 & $4.76 \times 10^{19}$ \\
    100M  & 8 & 8 & 1,024 & 4.5 days& H100 & $1.66 \times 10^{20}$ \\
    \bottomrule
  \end{tabular}
\end{table}

During training, promoters were truncated to a maximum length of 300 bp and randomly cropped (with a minimum size of 99 bp) as an augmentation to drive the learning of diverse features. Promoters were tokenized using overlapping 3-mers with a vocabulary built from `A', `C', `G', `T', and `N' (often found in assembled genome sequences to indicate gaps). A \texttt{<CLS>} token was prepended to all promoter sequences. Each of the four C3P models trained contains a transformer with rotary positional embedding \cite{su2021roformerenhancedtransformerrotary} as the promoter encoder. The only architectural differences between the models are changes in the number of transformer layers, attention heads per layer, and the hidden size $d_{promoter}$ of the promoter encoder. The output of the promoter encoder is the final layer embedding of the \texttt{<CLS>} token, which is then projected to $d_{projection}=256$ by the projection layer.

Protein representations from the frozen ESM2 (150M) encoder were pre-computed for the train and validation datasets (downloaded from \url{https://huggingface.co/facebook/esm2_t30_150M_UR50D}, MIT license). Proteins were truncated to a maximum tokenized length (using the ESM2 tokenizer) of 1,024 and their embeddings computed as the average of the final hidden layer representations from the ESM2 model, and then (during training) passed to the learned protein linear projection layer which reduces the dimensionality to $d_{projection}$. As an additional step to reduce compute requirements, for each dataset, proteins were clustered using MMSeqs2 \cite{steinegger_mmseqs2_2017} to 80\% identity and 80\% coverage, and representations only pre-computed for the representative proteins of each cluster. During training, proteins were mapped to their cluster and the embedding of their representative was retrieved. This reduced the number of unique proteins in the training dataset from 65 million to 26 million.

Table \ref{architecture-table} provides an overview of the variable parameters of the four C3P models trained, as well as their training compute requirements (FLOPS calculated as described in Appendix \ref{sec:scaling}). Each was trained for 10 epochs (875 million total examples). We used a batch size of 512 and the AdamW optimizer \cite{loshchilov2019adamw} with a learning rate of $1\times10^{-4}$, weight decay of 0.01, a cosine learning rate scheduler with 1000 warm-up steps, $\beta_1=0.9$ and $\beta_2=0.999$.

As promoter-protein pairs were randomly sampled to create training batches, and because proteins sharing an operon are paired with the same promoter sequence, it is possible a given promoter or protein will have multiple valid pairs in a batch. In this case, as only the sampled pair is considered a positive match, any others will be false negatives. While this is likely detrimental to model performance, we reasoned that with a batch size of 512 and our training dataset containing 87.5 million pairs from 35,878 genomes, such false negatives would occur infrequently and thus have a minor impact.

\section{Baseline Genome Language Models}
\label{sec:baselines}

Four gLMs were used as baselines to compare against the performance of C3P. Key characteristics of each model, as well as the manner in which we computed embeddings from them, are described below.

\paragraph{Evo (7B)} Evo \cite{nguyen_sequence_2024} is a long-context language model which was pretrained at single-nucleotide level through NTP on 80,000 prokaryotic genomes along with millions of predicted phage and plasmid sequences. We use the 7 billion parameter model which was trained on this dataset with a context-length of 8,192 (downloaded from \url{https://huggingface.co/togethercomputer/evo-1-8k-base}, apache-2.0 license). This was accomplished through the use of the StripedHyena \cite{stripedhyena} architecture, an efficient transformer alternative. In their evaluations the authors used the final layer embeddings of the Evo 7B model, averaged over the sequence length and having a dimensionality of 4,096. We extract embeddings of promoter sequences from Evo in the same manner. \textbf{Evo2 (7B)} \cite{brixi_genome_2026} uses a similar training regime and architecture to Evo (7B), but was trained on a significantly larger dataset (300B vs. 2.4T tokens) with sequences from all domains of life (downloaded from \url{https://huggingface.co/arcinstitute/evo2_7b}, apache-2.0 license). Rather than extracting embeddings from the final hidden layer, we extract Evo2 embeddings from layer 26 as that was the layer used by the authors in their evaluation. The layer 26 hidden representations were averaged over sequence length to have a dimensionality of 4,096.

\paragraph{PromoGen2 (149M)} PromoGen2 \cite{xia_design_2026} is a transformer-based gLM trained through NTP on a set of prokaryotic promoter sequences. In a manner very similar to our own for promoter extraction, 59 million promoters were extracted from 17,000 prokaryotic genomes by taking the non-coding regions 160 bp upstream of each start codon in each genome. Unlike our approach, the authors reduced this to 1.4 million training sequences by clustering by sequence similarity and only retaining cluster representatives with high predicted transcriptional strength. Similarly to Evo, the capacity of PromoGen2 to learn regulatory function was shown zero-shot through the correlation of the likelihood assigned by the model to promoters with matched genes expression levels. As the 149 million parameter model showed the strongest performance at this task (including outperforming Evo), we chose this model as our baseline (downloaded from \url{https://huggingface.co/jinyuan22/promogen2-base}, cc-by-nc-4.0
license). Embeddings from PromoGen2 were extracted by taking the average of the penultimate hidden layer representation over the sequence length (dimensionality of 640), as empirically this gave strong performance. Unlike all other models, promoters were truncated to a maximum length of 160 bp before being embedded with PromoGen2, as this was the length seen during its training.

\paragraph{gLM2 (650M)} gLM2 \cite{cornman_omg_2024} is a transformer-based gLM trained through mixed-modality MLM, where sequences are first preprocessed to be tokenized as nucleotides in non-coding regions and as amino acids within coding sequences (downloaded from \url{https://huggingface.co/tattabio/gLM2_650M}, apache-2.0 license). gLM2 was trained on 271 million contigs (containing 3.3 billion coding sequences and 2.8 billion intergenic regions) derived from metagenomic sequencing data (containing primarily prokaryotic and viral sequences). Using categorical Jacobian analysis \cite{zhang_protein_2024}, it was shown that gLM2 learned the boundaries of the sigma factor binding motifs in an \textit{E. coli} promoter. Promoter embeddings from gLM2 were also extracted by taking the average of the penultimate hidden layer representation over the sequence length (dimensionality of 1,280).

\section{Regulon/Sigmulon Prediction Evaluation}

\subsection{RegulonDB Evaluation}
\label{sec:regulon}

\begin{figure}[h]
    \centering
    \includegraphics[width=0.8\linewidth]{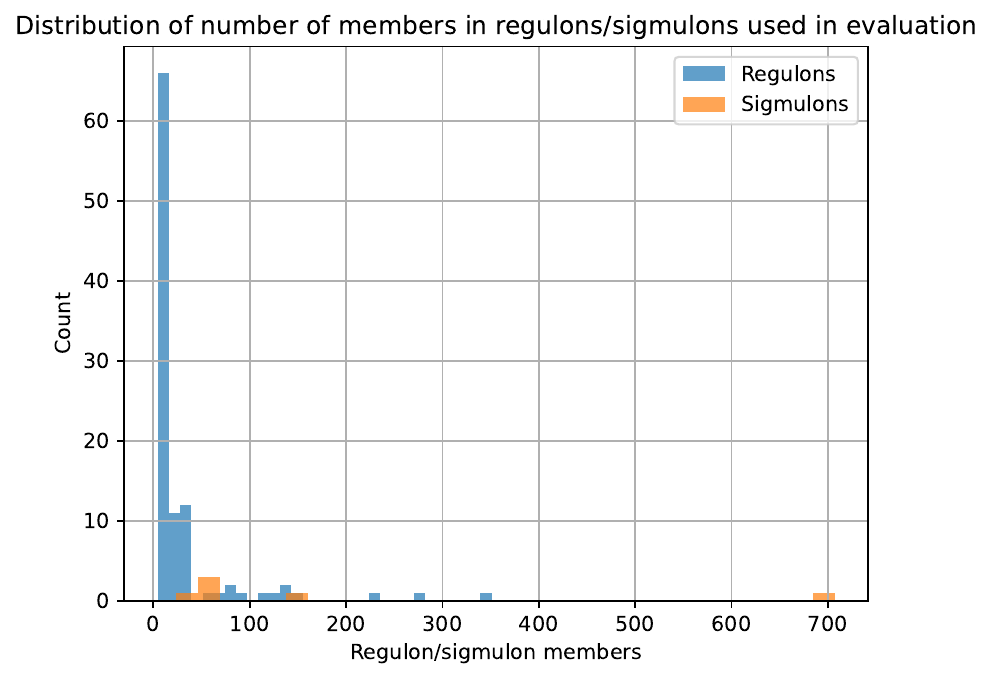}
    \caption{Distribution of the number of genes annotated for each of the 102 regulons and 6 sigmulons evaluated in our RegulonDB evaluation task.}
    \label{fig:distribution}
\end{figure}

Of the 2,902 genes in the \textit{E. coli} K-12 reference genome (GCF\_000005845.2) which we could extract promoter-protein pairs for, 2,102 were either standalone genes or the first member of a multi-gene operon. For this evaluation we do not include more than one gene per operon as, given members of an operon share a promoter sequence and are very likely to be co-regulated, this would inflate the performance of our approach in comparison with ESM2. RegulonDB \cite{salgado_regulondb_2024} (a database of \textit{E. coli} regulatory annotations freely available for academic use \url{https://regulondb.ccg.unam.mx/}) regulon and sigmulon annotations were downloaded on November 21, 2025 and November 27, 2025 respectively. Of the 2,102 genes, 1,340 could be associated with at least one regulon, and 889 with at least one sigmulon.

The frequency of different regulon and sigmulon classes varies widely (Figure \ref{fig:distribution}), with many having very few members. Given this imbalance, we chose to perform an unsupervised evaluation of the clustering of the embedding space separately for each class. Of the 275 unique regulons in RegulonDB we evaluated only on those with at least 5 members in our 2,102 genes, giving us 102 regulons and 6 sigmulons. We first performed binary $k=1$ k-nearest neighbours prediction of each class using embeddings from each method, extracted as previously described (Appendix \ref{sec:baselines}). Positives are promoter embeddings from genes annotated with a class and negatives are promoter embeddings from all genes in the 2,102 which do not have that class label (protein embeddings are used in the case of ESM2). Pairwise neighbour distances were calculated using cosine similarity.

Beyond measuring representation quality through binary classification performance, for each of our 102 regulon and 6 sigmulon classes we also measured the clustering of the embeddings beyond single nearest neighbours using mean Average Precision@R (mAP@R)\cite{musgrave_metric_2020}, an information retrieval metric. We can frame each gene in a regulon/sigmulon as a query, and measure our ability to retrieve other members of that regulon/sigmulon based on nearest neighbours in the embedding space.

In our binary setup where $R$ is the number of genes in a regulon/sigmulon minus one, let $i$ be a query gene within that regulon/sigmulon, $\text{mAP@R}_i$ for this gene is calculated as:

\begin{equation}
    \text{mAP@R}_i = \frac{1}{R} \sum^{R}_{j=1} P(j), \text{where } P(j) = 
    \begin{cases} 
        \text{precision at } j, & \text{if } j\text{-th nearest neighbour belongs} \\
                                & \text{to the same regulon/sigmulon} \\ 
        0,                      & \text{otherwise}
    \end{cases}
\end{equation}
The mAP@R for the regulon/sigmulon is then the average of the individual $\text{mAP@R}_i$ scores across all genes within that class (all queries). We report the macro-average of mAP@R over all regulons and sigmulons in Table \ref{gene-representation-metrics}. Average Precision is the \textit{de facto} retrieval metric used in vision tasks with imbalanced datasets (few positives and many negatives) \cite{ramzi_robust_2021}. We use mAP@R rather than mAP as it is less impacted by very distant positives. We also do not simply set an arbitrary $k$ and perform $\text{mAP@}k$ as regulons/sigmulons can significantly vary in size.

\subsection{Linear Probing Evaluation}
\label{sec:linear-probing}
\begin{table}[ht]
  \caption{Performance of C3P versus baseline gLMs and ESM2 at regulon classification through linear probing. Values represent the macro-average across all regulons with at least 30 members. Uncertainty represents 2$\times$ standard error of the mean. In brackets are trainable parameter counts.}
  \label{linear-probing-metrics}
  \small
  \centering
  \setlength{\tabcolsep}{3pt}
  \begin{tabular}{lcc}
    \toprule
    & \multicolumn{2}{c}{Regulons ($n=24$)}  \\
    \cmidrule(r){2-3} 
    Model & Precision & MCC \\
    \midrule
    \textbf{Ours} \\
    C3P (100M)  & \textbf{0.254$\pm.056$} & \textbf{0.193$\pm.047$}  \\
    \midrule
    \textbf{Our pLM teacher} \\
    ESM2 (150M)   & 0.087$\pm$.026 & 0.093$\pm$.028  \\
    \midrule
    \textbf{Mixed-modality gLM} \\
    gLM2 (650M)   & 0.090$\pm$.030 & 0.058$\pm$.024 \\    
    \midrule
    \textbf{DNA gLMs} \\
    Evo2 (7B)   & 0.100$\pm$.028 & 0.091$\pm$.027 \\
    Evo (7B)    & 0.057$\pm$.022 & 0.046$\pm$.019 \\
    PromoGen2 (149M) & 0.069$\pm$.021 & 0.039$\pm$.017  \\   
    \bottomrule
  \end{tabular}
\end{table}

In addition to comparing 1-NN regulon prediction performance using embeddings from each method, we also compared the performance of linear models. Using the default parameters (except setting max\_iter=200 and class\_weight='balanced') of the \texttt{LogisticRegression} module from the scikit-learn Python package \cite{scikit-learn}, models were fit through stratified 5-fold cross-validation for predicting membership in each regulon with at least 30 members (n=24). Table \ref{linear-probing-metrics} contains the results of this evaluation, averaged across the 24 regulons. A similar performance trend to the 1-NN regulon prediction performance can be seen, with C3P widely outperforming other models.

\subsection{DBTBS Evaluation}
\label{sec:dbtbs}
\begin{table}[ht]
  \caption{Performance of C3P versus baseline gLMs and ESM2 at \textit{B. subtilis} regulon/sigmulon classification. Values represent the macro-average across all regulons with at least 5 members and all sigmulons. Uncertainty represents 2$\times$ standard error of the mean. 1-nn precision (Prec.), 1-nn Matthews correlation coefficient (MCC), and mean Average Precision@R (mAP@R)\cite{musgrave_metric_2020} shown. In brackets are trainable parameter counts.}
  
  \label{bsub-gene-representation-metrics}
  \small
  \centering
  \setlength{\tabcolsep}{3pt}
  \begin{tabular}{lcccccc}
    \toprule
    & \multicolumn{3}{c}{Regulons ($n=21$)} & \multicolumn{3}{c}{Sigmulons ($n=11$)} \\
    \cmidrule(r){2-4} \cmidrule(l){5-7}
    Model & Prec. (1-nn) & MCC (1-nn) & mAP@R  & Prec. (1-nn) & MCC (1-nn) & mAP@R \\
    \midrule
    \textbf{Ours} \\
    C3P (100M)  & \textbf{0.247$\pm.113$} & \textbf{0.260$\pm.119$} & \textbf{0.119$\pm.072$} & \textbf{0.384$\pm.134$} & \textbf{0.367$\pm.144$} & \textbf{0.200}$\pm.147$ \\
    \midrule
    \textbf{Our pLM teacher} \\
    ESM2 (150M)   & 0.109$\pm$.043 & 0.110$\pm$.048 & 0.030$\pm$.018 & 0.132$\pm$.086 & 0.108$\pm$.080 & 0.018$\pm$.019 \\
    \midrule
    \textbf{Mixed-modality gLM} \\
    gLM2 (650M)   & 0.013$\pm$.018 & 0.005$\pm$.016 & 0.004$\pm$.002& 0.027$\pm$.032 & 0.008$\pm$.017& 0.008$\pm$.006 \\    
    \midrule
    \textbf{DNA gLMs} \\
    Evo2 (7B)        & 0.036$\pm$.035 & 0.035$\pm$.041 & 0.013$\pm$.012 & 0.178$\pm$.178 & 0.098$\pm$.080 & 0.014$\pm$.009 \\
    Evo (7B)         & 0.004$\pm$.007 & -0.003$\pm$.006 & 0.001$\pm$.001 & 0.015$\pm$.022 & -0.006$\pm$.010 & 0.003$\pm$.003 \\
    PromoGen2 (149M) & 0.019$\pm$.026 & 0.009$\pm$.021 & 0.004$\pm$.003 & 0.028$\pm$.026 & 0.006$\pm$.018 & 0.006$\pm$.004 \\    
    \bottomrule
  \end{tabular}
\end{table}

A curated list of transcription and sigma factors and the genes regulated by them was pulled from the DBTBS database \cite{sierro_dbtbs_2008} on July 30 2026, representing the regulons and sigmulons of \textit{B. subtilis}. After these annotations were matched with the 2,046 standalone/first operon member genes in the \textit{B. subtilis} reference genome (GCF\_000009045.1), 21 regulons and 11 sigmulons had at least 5 members. As was done for \textit{E. coli}, membership in each of these regulons and sigmulons was predicted separately through 1-NN binary classification. Table \ref{bsub-gene-representation-metrics} contains the results of this evaluation, which show a similar trend to the performance trend seen for 1-NN \textit{E. coli} regulon and sigmulon prediction, with C3P widely outperforming other models. This evaluation validates that curated regulatory annotations can be well-predicted with C3P representations beyond \textit{E. coli}.

\newpage

\section{Zero-Shot Co-Regulated Gene Retrieval Evaluation}
\label{sec:imodulondb}

\begin{table}[h]
\centering
\caption{Overview of the iModulonDB datasets. Only standalone genes and the first genes in each multi-gene operon are included. \# iModulons indicates number of iModulons which could be associated with at least one gene. \# Genes indicates number of genes with at least one iModulon annotated. Species in train indicates whether or not genomes from that species were used in the train dataset. }
\label{tab:imodulons_summary}
\begin{tabular}{lccccc}
\toprule
 &  &  & &  & Species \\
Study & Species & Accession & \# iModulons & \# Genes & in train \\ \midrule
precise139 & \textit{A. baumannii} &GCF\_000963815.1 & 47 & 439 & Yes \\
modulome & \textit{B. diazoefficiens}&GCF\_000011365.1  & 62 & 1,122 & Yes \\
nicolas & \textit{B. subtilis} & GCF\_000009045.1& 71 & 627 & Yes \\
modulome & \textit{B. subtilis}&GCF\_000009045.1  & 71 & 844 & Yes \\
modulome263 & \textit{C. glutamicum}&GCF\_000382905.1 & 72 & 447 & Yes \\
precise\_mg1655 & \textit{E. coli}&GCF\_000005845.2 & 106 & 719 & No \\
precise815 & \textit{E. coli}& GCF\_000005845.2& 196 & 1,190 & No \\
precise278 & \textit{E. coli}& GCF\_000005845.2& 92 & 742 & No \\
modulome & \textit{E. coli}& GCF\_000005845.2& 186 & 1,229 & No \\
precise1k & \textit{E. coli}&GCF\_000005845.2 & 190 & 1,252 & No \\
precise101 & \textit{L. plantarum}& GCF\_000203855.3& 41 & 184 & Yes \\
lactoprecise & \textit{L. reuteri}&GCF\_020785475.1 & 35 & 273 & Yes \\
modulome & \textit{M. tuberculosis}& GCF\_000195955.2& 69 & 459 & Yes \\
precise411 & \textit{P. aeruginosa}& GCF\_000006765.1& 107 & 1,136 & Yes \\
precise364 & \textit{P. aeruginosa}& GCF\_000006765.1& 101 & 1,136 & Yes \\
precise321 & \textit{P. putida}&GCF\_000007565.2 & 77 & 604 & Yes \\
staph\_precise165 & \textit{S. aureus}& GCF\_000017085.1& 74 & 578 & Yes \\
staph\_precise108 & \textit{S. aureus}&GCF\_000017085.1 & 29 & 339 & Yes \\
modulome478 & \textit{S. coelicolor}& GCF\_000203835.1& 116 & 2,701 & No \\
elprecise300 & \textit{S. elongatus}&GCF\_000012525.1 & 55 & 468 & Yes \\
core & \textit{S. enterica}&GCF\_000006945.2 & 111 & 739 & No \\
precise718 & \textit{S. pneumoniae}& GCF\_000006885.1& 57 & 245 & No \\
modulome & \textit{S. pyogenes}& GCF\_001021955.1& 42 & 243 & No \\
precise108 & \textit{V. natriegens}&GCF\_001456255.1 & 61 & 625 & Yes \\
precise104 & \textit{V. natriegens}& GCF\_001456255.1& 40 & 453 & Yes \\
\bottomrule
\end{tabular}
\end{table}

\begin{table}[h]
\centering
\caption{Zero-shot co-regulated gene retrieval performance of C3P (100M) on each of the 25 iModulonDB datasets. Each accuracy treated as binomial proportion, uncertainty represents 2$\times$standard error of each binomial proportion.}
\label{tab:imodulons_performance}
\begin{tabular}{lccc}
\toprule
Study & Species & Top-1 full match accuracy & Top-1 any match accuracy \\ \midrule
precise139 & \textit{A. baumannii} & 0.280 $\pm$ 0.043 & 0.408 $\pm$ 0.047 \\
modulome & \textit{B. diazoefficiens} & 0.332 $\pm$ 0.028 & 0.489 $\pm$ 0.030 \\
nicolas & \textit{B. subtilis} & 0.458 $\pm$ 0.040 & 0.625 $\pm$ 0.039 \\
modulome & \textit{B. subtilis} & 0.220 $\pm$ 0.029 & 0.469 $\pm$ 0.034 \\
modulome263 & \textit{C. glutamicum} & 0.101 $\pm$ 0.028 & 0.159 $\pm$ 0.035 \\
precise\_mg1655 & \textit{E. coli} & 0.250 $\pm$ 0.032 & 0.399 $\pm$ 0.037 \\
precise815 & \textit{E. coli} & 0.108 $\pm$ 0.018 & 0.304 $\pm$ 0.027 \\
precise278 & \textit{E. coli} & 0.217 $\pm$ 0.030 & 0.357 $\pm$ 0.035 \\
modulome & \textit{E. coli} & 0.142 $\pm$ 0.020 & 0.337 $\pm$ 0.027 \\
precise1k & \textit{E. coli} & 0.108 $\pm$ 0.018 & 0.313 $\pm$ 0.026 \\
precise101 & \textit{L. plantarum} & 0.228 $\pm$ 0.062 & 0.261 $\pm$ 0.065 \\
lactoprecise & \textit{L. reuteri} & 0.220 $\pm$ 0.050 & 0.385 $\pm$ 0.059 \\
modulome & \textit{M. tuberculosis} & 0.303 $\pm$ 0.043 & 0.423 $\pm$ 0.046 \\
precise411 & \textit{P. aeruginosa} & 0.188 $\pm$ 0.023 & 0.430 $\pm$ 0.029 \\
precise364 & \textit{P. aeruginosa} & 0.202 $\pm$ 0.024 & 0.416 $\pm$ 0.029 \\
precise321 & \textit{P. putida} & 0.419 $\pm$ 0.040 & 0.480 $\pm$ 0.041 \\
staph\_precise165 & \textit{S. aureus} & 0.230 $\pm$ 0.035 & 0.420 $\pm$ 0.041 \\
staph\_precise108 & \textit{S. aureus} & 0.419 $\pm$ 0.054 & 0.552 $\pm$ 0.054 \\
modulome478 & \textit{S. coelicolor} & 0.075 $\pm$ 0.010 & 0.342 $\pm$ 0.018 \\
elprecise300 & \textit{S. elongatus} & 0.124 $\pm$ 0.030 & 0.325 $\pm$ 0.043 \\
core & \textit{S. enterica} & 0.237 $\pm$ 0.031 & 0.363 $\pm$ 0.035 \\
precise718 & \textit{S. pneumoniae} & 0.188 $\pm$ 0.050 & 0.314 $\pm$ 0.059 \\
modulome & \textit{S. pyogenes} & 0.119 $\pm$ 0.042 & 0.309 $\pm$ 0.059 \\
precise108 & \textit{V. natriegens} & 0.306 $\pm$ 0.037 & 0.410 $\pm$ 0.039 \\
precise104 & \textit{V. natriegens} & 0.435 $\pm$ 0.047 & 0.503 $\pm$ 0.047 \\
\bottomrule
\end{tabular}
\end{table}

Bottom-up experimental determination of gene co-regulation (as in RegulonDB \cite{salgado_regulondb_2024}) is only feasible for a handful of model organisms. Instead, independently regulated sets of genes (termed iModulons as they represent an independently modulated signal) can be inferred top-down by applying Independent Component Analysis (ICA) to collections of gene expression data gathered when various perturbations are applied to an organism. iModulonDB \cite{rychel_imodulondb_2021} is a database of ICA inferred iModulons from expression perturbation datasets from many bacterial species (freely available for academic use \url{https://imodulondb.org/}). From iModulonDB we selected the 25/28 datasets coming from 17/20 species for which we could retrieve annotated genomes from NCBI with gene IDs matching those in iModulonDB. Table \ref{tab:imodulons_summary} gives a summary of each dataset. 9/25 (5/17 species) were sourced from studies on species not included in our training dataset. Promoter-protein pairs were extracted from each genome as previously described (Appendix \ref{sec:training_data}). Table \ref{tab:imodulons_performance} gives the per-dataset performance of C3P.

We introduced the zero-shot co-regulated gene retrieval task described in section \ref{sec:task2} for predicting this data, using the iModulon annotations as indicators of what genes in a genome are co-regulated. This was done because while regulons/sigmulons indicate genes regulated by specific transcription/sigma factors, individual iModulons, though known to correlate with regulons/sigmulons \cite{sastry_escherichia_2019} do not have inherent regulatory meaning (although post-hoc analyses can assign probable regulators). This is exemplified by the variable number of iModulons contained in the 5 iModulonDB datasets for \textit{E. coli} (Table \ref{tab:imodulons_summary}), likely due to differences in the perturbation experiments performed in each study. Because of this, predicting individual iModulon membership, as was done for regulons and sigmulons in section \ref{sec:task1}, may not be informative of the regulatory features a model learns to capture.

\begin{table}[ht]
  \caption{Performance of C3P versus baseline gLMs and ESM2 at zero-shot co-regulated gene retrieval using the \textit{E. coli} regulon and sigmulon annotations in RegulonDB. Accuracy treated as a binomial proportion, uncertainty represents 2$\times$standard error of the binomial proportion.}
  \label{zero-shot-table-regulondb}
  \small
  \centering
  \setlength{\tabcolsep}{3pt}
  \begin{tabular}{lcccc}
    \toprule
    & \multicolumn{2}{c}{Regulons} & \multicolumn{2}{c}{Sigmulons} \\
    \cmidrule(r){2-3} \cmidrule(l){4-5}
          & Top-1 Full     & Top-1 Any      & Top-1 Full     & Top-1 Any \\
    Model & Match Accuracy & Match Accuracy & Match Accuracy & Match Accuracy \\
    \midrule
    \textbf{Ours} \\
    C3P (100M)       & \textbf{0.057$\pm.013$} & \textbf{0.438$\pm.027$} & \textbf{0.524$\pm.033$} & \textbf{0.758$\pm.029$} \\
    \midrule
    \textbf{Our pLM teacher} \\
    ESM2 (150M)      & 0.026$\pm.009$          & 0.309$\pm.025$          & 0.445$\pm.033$          & 0.691$\pm.031$          \\
    \midrule
    \textbf{Mixed-modality gLM} \\
    gLM2 (650M)      & 0.028$\pm.009$          & 0.271$\pm.024$          & 0.450$\pm.033$          & 0.692$\pm.031$          \\  
    \midrule
    \textbf{DNA gLMs} \\
    Evo2 (7B)        & 0.031$\pm.010$          & 0.240$\pm.023$          & 0.444$\pm.033$          & 0.685$\pm.031$          \\
    Evo (7B)         & 0.019$\pm.007$          & 0.196$\pm.022$          & 0.404$\pm.033$          & 0.642$\pm.032$          \\
    PromoGen2 (149M) & 0.028$\pm.009$          & 0.242$\pm.023$          & 0.441$\pm.033$          & 0.669$\pm.032$          \\   
    \bottomrule
  \end{tabular}
\end{table}

While the above reasoning explains why we did not use the evaluation method applied in section \ref{sec:task1}, it does not preclude the application of the zero-shot co-regulated gene retrieval evaluation to the RegulonDB regulon/sigmulon data, as this data also indicates co-regulated genes. Table \ref{zero-shot-table-regulondb} contains the results of performing such an evaluation. C3P again widely outperforms all baseline models.

The UMAP plots seen in Figure \ref{fig:umap} were created with embeddings for all 2,046 genes (standalone genes and the first members of multi-gene operons)  in the \textit{B. subtilis} genome using the umap-learn package \cite{mcinnes_umap_2020} with distance=`cosine', random\_state=0 for the C3P plot, random\_state=1 for the ESM2 plot, and default settings otherwise.

\newpage
\section{Random Baseline Comparison}
\label{sec:randombaseline}

Randomly initialized versions of each baseline gLM as well as C3P were compared against the trained performance of each model (Section \ref{sec:task3}), with embeddings being extracted in the same manner as the trained models. Weights of the randomly initialized Evo2 parameters were constrained to be normally distributed with a mean of 0 and standard deviation of 0.005, as failure to constrain the weights resulted in embeddings containing NaN.

\newpage
\section{Scaling Evaluation}
\label{sec:scaling}

\begin{figure}[h]
    \centering
    \includegraphics[width=0.7\linewidth]{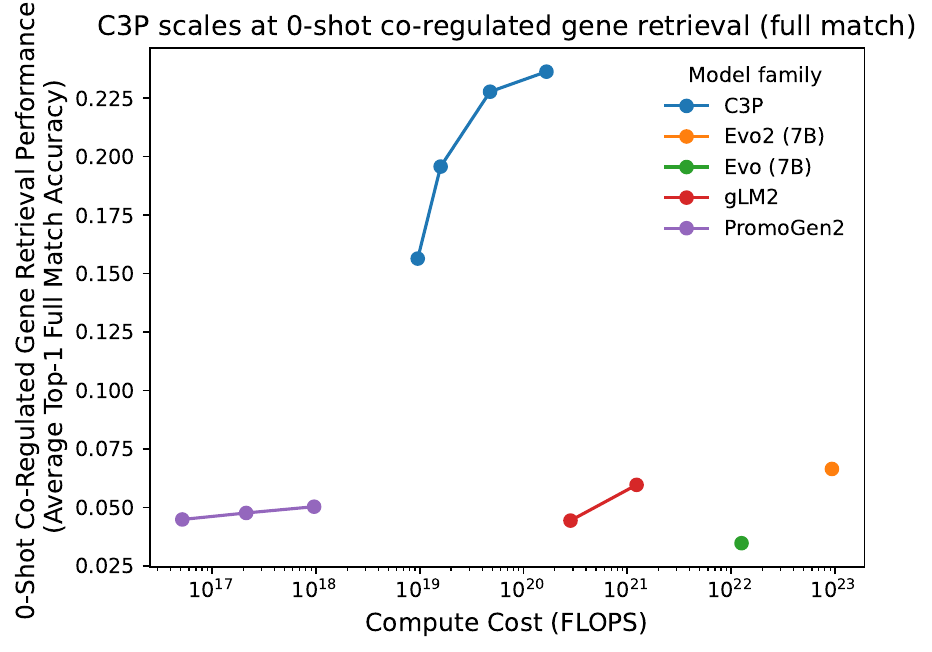}
    \caption{Scaling of C3P and each gLM baseline. Average zero-shot co-regulated gene retrieval top-1 full match accuracy across the 25 iModulonDB datasets is compared to training FLOPS.}
    \label{fig:flops2}
\end{figure}

\begin{table}[h]
\centering
\caption{Values used for pretraining floating point operations (FLOPS) calculation for all models. All values are rounded approximations. }
\label{tab:scaling}
\begin{tabular}{lccccc}
\toprule
Model & Parameters & Tokens & Epochs & Pre-embedding FLOPS & Total FLOPS \\ \midrule
C3P & 100M & 300 $\times$ 88M& 10 &8.0 $\times$ 10$^{18}$& 1.66 $\times$ 10$^{20}$\\
C3P & 25M & 300 $\times$ 88M& 10 &8.0 $\times$ 10$^{18}$& 4.76 $\times$ 10$^{19}$\\
C3P & 5M & 300 $\times$ 88M& 10 &8.0 $\times$ 10$^{18}$& 1.59 $\times$ 10$^{19}$\\
C3P & 1M & 300 $\times$ 88M& 10 &8.0 $\times$ 10$^{18}$& 9.57 $\times$ 10$^{18}$\\
PromoGen2 & 149M & 256 $\times$ 1.4M& 3 & - & 9.61 $\times$ 10$^{17}$\\
PromoGen2 & 33M & 256 $\times$ 1.4M& 3 & - & 2.13 $\times$ 10$^{17}$\\
PromoGen2 & 8M & 256 $\times$ 1.4M& 3 & - & 5.16 $\times$ 10$^{16}$\\
gLM2 & 650M & 315B & 1 & - & 1.23 $\times$ 10$^{21}$\\
gLM2 & 150M & 315B & 1 & - & 2.84 $\times$ 10$^{20}$\\
Evo & 7B &  300B & 1 & - & 1.26 $\times$ 10$^{22}$\\
Evo2 & 6.5B & 2.4T & 1 & - & 9.36 $\times$ 10$^{22}$\\
\bottomrule
\end{tabular}
\end{table}

Pretraining floating point operations (FLOPS) were calculated according to the common heuristic FLOPS $= 6 \times N \times D$, where $N$ represents the number of model parameters and $D$ the number of tokens in the training dataset \cite{kaplan2020scalinglawsneurallanguage}. For the C3P models, pre-embedding FLOPS are also calculated as the forward pass FLOPS of pre-computing 26M protein embeddings with the ESM2 (150M) model (with a max length of 1,024 amino acids) as $2 \times 1{,}024 \times 26\text{M} \times 150\text{M} = 8.0\times10^{18}$. For token counts of the C3P models, the maximum promoter length (300 bp) multiplied by the number of training examples was used. During PromoGen2 training, sequences were padded to a length 256 bp, and this value multiplied by the number of training examples was used for the token count.

Table \ref{tab:scaling} shows the total FLOPS of each gLM and C3P, as well as the values used in the calculations. For models trained for more than one epoch, token counts are multiplied by the number of epochs. Evo2 (40B) was not included in this analysis due to a lack of sufficient computational resources. As the true parameter count of Evo2 7B is 6.5B, we use this value for FLOPS calculation. This is also true for Evo, but as FLOPS was previously reported for Evo using 7B as the parameter count, we report that value here.

Figure \ref{fig:flops} and Figure \ref{fig:flops2} show the scaling of regulon classification and zero-shot co-regulated gene retrieval performance for C3P and each gLM baseline.

\newpage
\section{Zero-shot Binding Site Prediction Evaluation}
\label{sec:segmentation}

The locations of all transcription factor and sigma factor binding sites annotated in the \textit{E. coli} K-12 reference genome (GCF\_000005845.2) were downloaded from RegulonDB on July 15 2026. From the 2,102 promoter-protein pairs extracted using the previously described approach (standalone genes and first members of multi-gene operons only, see Appendix \ref{sec:training_data}), the first 100 pairs with a promoter overlapping an annotated binding site (all sigma box sites, only "strong" or "confirmed" confidence level transcription factor binding sites) on the same strand, within the 100 bp upstream of start codon, were selected. The selected promoters were then truncated to 100 bp.

Binding sites were predicted using per-nucleotide importance scores determined through \textit{in-silico} mutagenesis. To perform \textit{in-silico} mutagenesis with the C3P model, for each of these 100 promoters and their corresponding proteins, we calculate the absolute change in the promoter-protein cosine similarity caused by mutating the wild-type promoter sequence at each position, averaging over the three mutations possible at each position to get $mutation\_score_{i}$ for $i \in [1,. .. , 100]$. For a wild-type promoter of length $L=100$, $wt \in \{A, C, T, G\}^{L}$, the promoter-protein cosine similarity is $S_{wt}$. When $wt$ is mutated at position $i$ to have base $a  \in \{A,C,T,G\}, a \neq wt_i$, promoter-protein cosine similarity becomes $S_{i, a}$. We calcualte $mutation\_score_{i}$ for each position in each promoter as follows:

\begin{equation}
    mutation\_score_{i}=( \sum_{a \in \{A,C,T,G\},a\neq wt_i}|S_{wt}-S_{i, a}| )/3
\end{equation}

Using the previously described same strand sigma factor and transcription factor binding site locations as the binary prediction target $binding\_sites= \{0,1\}^L$, for each sequence we calculate area under the ROC curve $AUROC(mutation\_score, binding\_sites)$ and area under the precision-recall curve
$AUPRC(mutation\_score, binding\_sites)$.

We compared average C3P performance against each gLM baseline, where $\Delta likelihood\_score_{i, a}$ is used in place of $S_{wt} - S_{i, a}$. Evo, Evo2, and PromoGen2 are autoregressive models, and as such $\Delta likelihood\_score_{i, a}$ represents the absolute change in the sequence log-likelihood score compared to wild-type when position $i$ of the promoter is mutated to base $a$. For Evo and Evo2, the DNA sequence of the promoter and corresponding gene were concatenated before being scored by the model. This allows for the likelihood of the sequence to be calculated in the context of the protein the promoter regulates. For PromoGen2 only the promoter sequence was provided as it was not trained on protein-coding DNA.

As gLM2 was trained through masked-language modeling, exact sequence likelihood cannot be calculated as for the autoregressive gLMs. Therefore, for gLM2, $\Delta likelihood\_score_{i, a}$ represents the change in sequence pseudolikelihood when position $i$ is mutated to base $a$, calculated as described in prior work \cite{hou_understanding_2026}. As gLM2 is a mixed-modality gLM, the promoter sequence (as nucleotides) and the protein sequence (as amino acids) were concatenated before scoring. Due to computational constraints, only the first 1024 amino acids of each protein were included.

\newpage
\section{Promoter Strength Evaluation}
\label{sec:evotask}

\begin{figure}[h]
    \centering
    \includegraphics[width=1.0\linewidth]{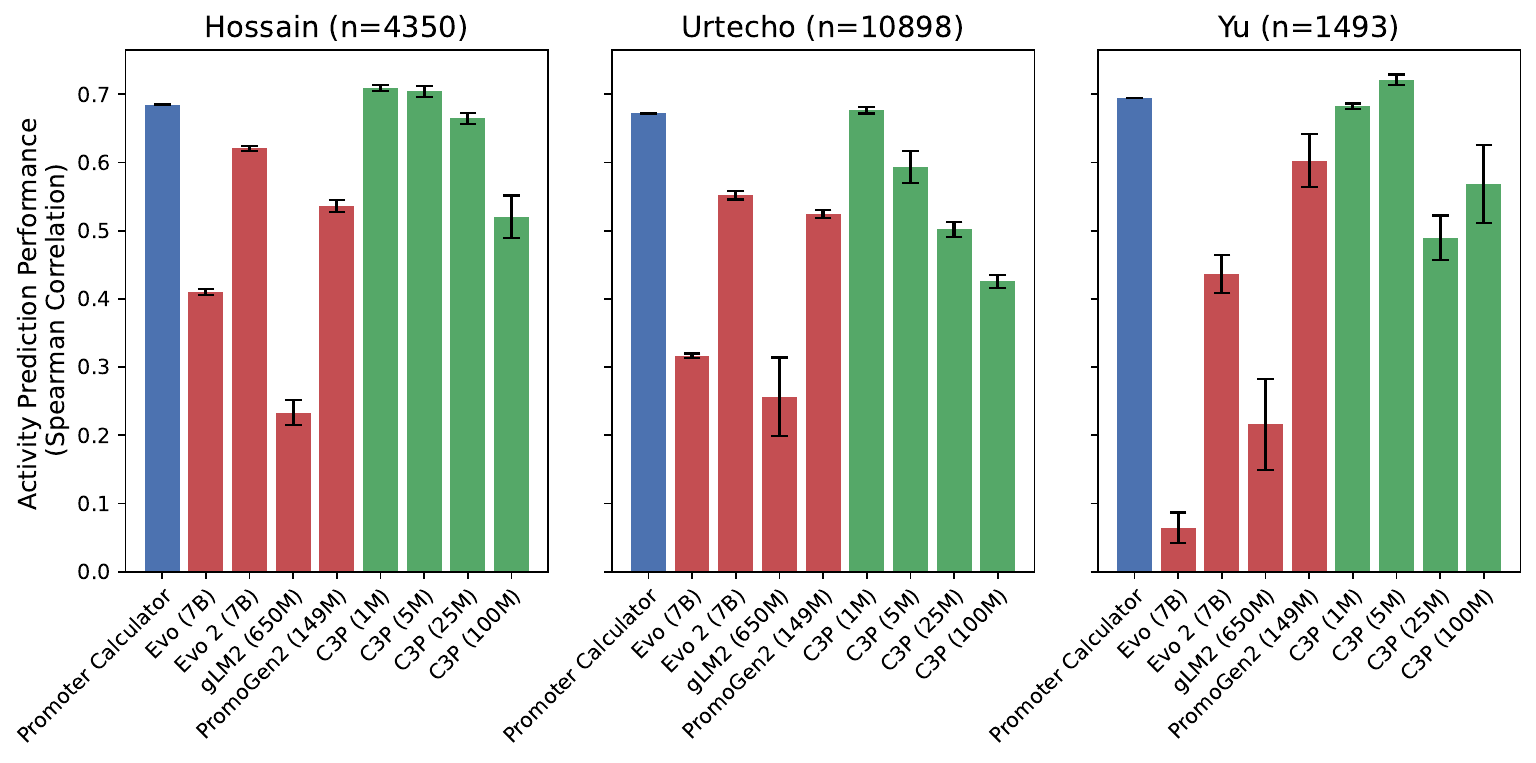}
    \caption{Spearman's correlation of predictions from ridge regression models trained with embeddings from each gLM baseline (red) and C3P model (green) on three \textit{in vivo} promoter activity datasets. Performance of a state-of-the-art supervised biophysical model (Promoter Calculator \cite{lafleur_automated_2022}, in blue) is included. Models were trained on activity measurements from an \textit{in vitro} dataset 5 times with different 80\% random splits. The mean performance is reported here, with uncertainty being 2$\times$standard error of the mean.}
    \label{fig:evoperformance}
\end{figure}

\begin{table}[h]
\centering
\caption{Spearman's correlation of predictions from ridge regression models trained with embeddings from each gLM baseline and C3P model on three \textit{in vivo} promoter activity datasets. Performance of a state-of-the-art supervised biophysical model (Promoter Calculator \cite{lafleur_automated_2022}) is included. Models were trained on activity measurements from an \textit{in vitro} dataset 5 times with different 80\% random splits. The mean performance is reported here, with uncertainty being 2$\times$standard error of the mean. }
\label{tab:promoter_strength}
\begin{tabular}{lccc}
\toprule
Model & Hossain (n=4350) & Urtecho (n=10898) & Yu (n=1493) \\
\midrule
C3P (1M) & \textbf{0.709}  $\pm$ 0.004 & \textbf{0.677}  $\pm$ 0.005 & 0.682  $\pm$ 0.004 \\
C3P (5M) & 0.704  $\pm$ 0.008 & 0.594  $\pm$ 0.023 & \textbf{0.721}  $\pm$ 0.008 \\
C3P (25M) & 0.664  $\pm$ 0.008 & 0.502  $\pm$ 0.011 & 0.489  $\pm$ 0.033 \\
C3P (100M) & 0.521  $\pm$ 0.031 & 0.426  $\pm$ 0.009 & 0.568  $\pm$ 0.057 \\
Evo2 (7B) & 0.621  $\pm$ 0.004 & 0.552  $\pm$ 0.006 & 0.436  $\pm$ 0.028 \\
Evo (7B) & 0.410  $\pm$ 0.004 & 0.317  $\pm$ 0.003 & 0.065  $\pm$ 0.022 \\
gLM2 (650M) & 0.233  $\pm$ 0.019 & 0.257  $\pm$ 0.058 & 0.216  $\pm$ 0.067 \\
PromoGen2 (149M) & 0.536  $\pm$ 0.009 & 0.525  $\pm$ 0.005 & 0.603  $\pm$ 0.039 \\
Promoter Calculator & 0.685  $\pm$ 0.000 & 0.672  $\pm$ 0.000 & 0.694  $\pm$ 0.000 \\
\bottomrule
\end{tabular}
\end{table}

The focus of our evaluations was to determine the extent that the learned representations from C3P and each gLM baseline distinguish differentially regulated bacterial promoters. However, we also evaluated the capacity for linear models trained on these representations to predict promoter activity levels, as this task has previously been used for evaluation of whether gLMs learn bacterial regulatory sequence function \cite{li_unveiling_2025, nguyen_sequence_2024, xia_design_2026}.

Evo \cite{nguyen_sequence_2024} is one gLM which was previously evaluated with this task. One approach used in this evaluation was to train a ridge regression model for predicting activity levels from promoter sequence embeddings extracted from Evo. They trained on a dataset containing 5,193 synthetic promoters paired with expression measurements from an \textit{in vitro} assay \cite{lafleur_automated_2022}, then evaluated the performance of their model on multiple datasets of \textit{in vivo} experiments (Hossain et al. \cite{hossain_automated_2020}, Urtecho et al. \cite{urtecho_systematic_2019}, Yu et al. \cite{yu_multiplexed_2021}) where the strength of thousands of synthetic promoters was measured by the expression levels of a reporter gene in \textit{E. coli}.

We followed the approach of Evo for our evaluation. Background sequences were first removed from the promoters of each of the four datasets by finding the subsequence within each with the strongest predicted forward strand activity according to Promoter Calculator \cite{lafleur_automated_2022}, a state-of-the-art supervised biophysical model. Using the default parameters of the \texttt{RidgeCV} module from the scikit-learn Python package \cite{scikit-learn}, models were fit on the training dataset using embeddings of the promoters from each gLM baseline as well as the C3P model. Each model was then evaluated on each of the three \textit{in vivo} datasets. 

Figure \ref{fig:evoperformance} and Table \ref{tab:promoter_strength} show the performance of each method, as well as Promoter Calculator. Unlike other evaluations, the 1M and 5M parameter C3P models show best performance, each exceeding the supervised baseline on 2/3 datasets, and outperforming all gLM baselines on 3/3 datasets. The best performing gLM varies by dataset between Evo2 and PromoGen2.

\begin{figure}[h]
    \centering
    \includegraphics[width=1.0\linewidth]{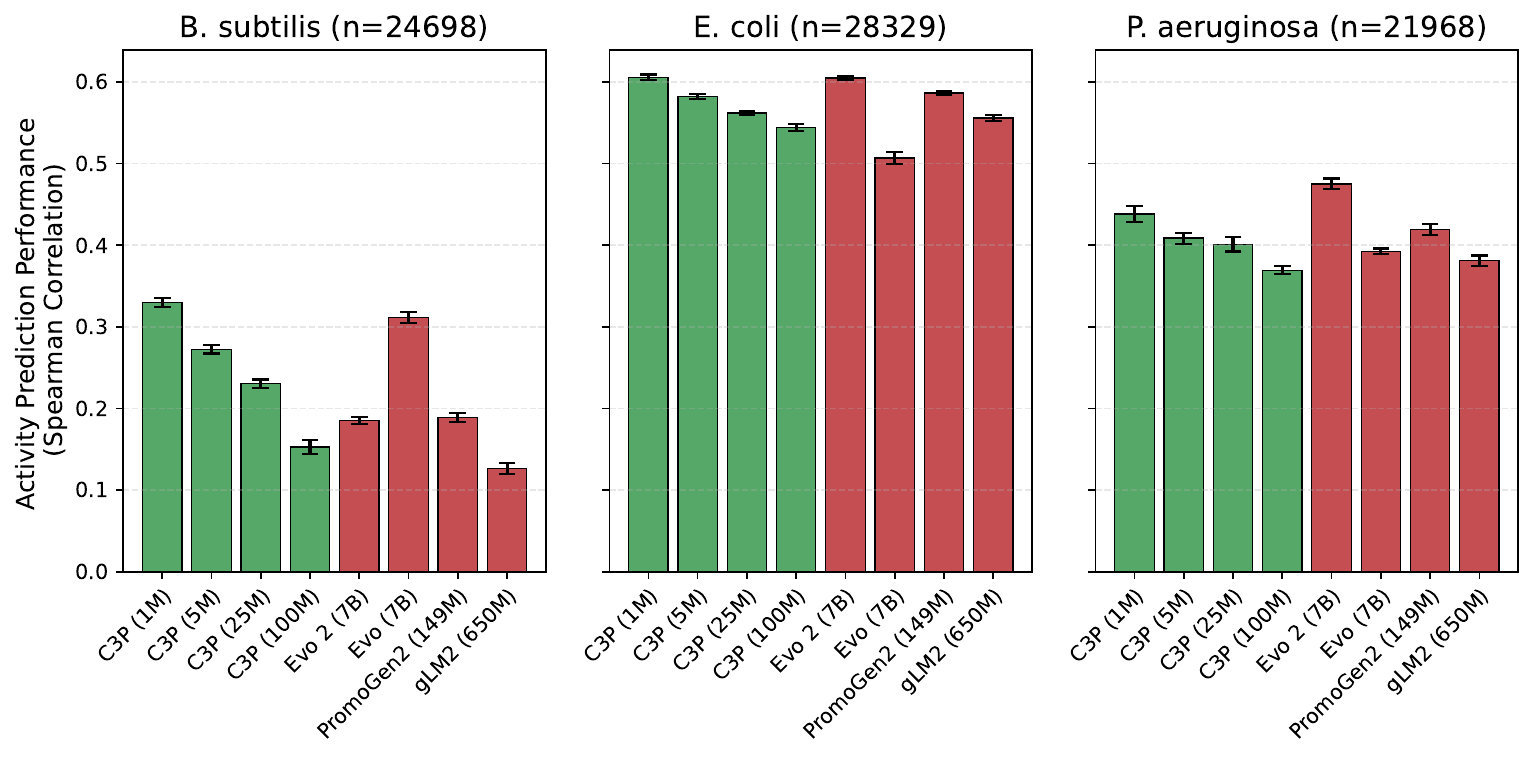}
    \caption{Spearman's correlation of predictions from ridge regression models trained with embeddings from each gLM baseline (red) and C3P model (green) on wild-type promoter activity levels in three model species. Models were trained 5 times with different 80\% random splits. The mean performance is reported here, with uncertainty being 2$\times$standard error of the mean.}
    \label{fig:evojohnsperformance}
\end{figure}

To confirm the trend that larger C3P models show decreased performance on this task, we repeated this evaluation on a dataset measuring the \textit{in vivo}  activity levels of a diverse collection of 29,248 wild-type promoters (each 165 bp) in three model species (\textit{E. coli}, \textit{B. subtilis}, \textit{P. aeruginosa}) \cite{johns_metagenomic_2018}. 

For each species, promoters with non-numeric normalized transcription values were removed. Linear models for predicting promoter activity were then fit using embeddings from each method using the default parameters of the \texttt{RidgeCV} module from the scikit-learn Python package \cite{scikit-learn}. Figure \ref{fig:evojohnsperformance} shows the performance of each method on this dataset. As before, smaller C3P models show stronger performance, and baseline gLM performance varies by species. Unlike on the datasets used in the Evo evaluation, C3P does not achieve clear state-of-the-art performance.

\newpage
\section*{NeurIPS Paper Checklist}

\begin{enumerate}

\item {\bf Claims}
    \item[] Question: Do the main claims made in the abstract and introduction accurately reflect the paper's contributions and scope?
    \item[] Answer: \answerYes{} 
    \item[] Justification: All claims made in the abstract and introduction are made with appropriate qualifications and substantiated by the results of the performed experiments
    \item[] Guidelines:
    \begin{itemize}
        \item The answer \answerNA{} means that the abstract and introduction do not include the claims made in the paper.
        \item The abstract and/or introduction should clearly state the claims made, including the contributions made in the paper and important assumptions and limitations. A \answerNo{} or \answerNA{} answer to this question will not be perceived well by the reviewers. 
        \item The claims made should match theoretical and experimental results, and reflect how much the results can be expected to generalize to other settings. 
        \item It is fine to include aspirational goals as motivation as long as it is clear that these goals are not attained by the paper. 
    \end{itemize}

\item {\bf Limitations}
    \item[] Question: Does the paper discuss the limitations of the work performed by the authors?
    \item[] Answer: \answerYes{} 
    \item[] Justification: Limitations are discussed in the discussion section. This includes our lack of interpretability analysis and the impact of our use of a pre-trained model in comparing our model efficiency to existing approaches.
    \item[] Guidelines:
    \begin{itemize}
        \item The answer \answerNA{} means that the paper has no limitation while the answer \answerNo{} means that the paper has limitations, but those are not discussed in the paper. 
        \item The authors are encouraged to create a separate ``Limitations'' section in their paper.
        \item The paper should point out any strong assumptions and how robust the results are to violations of these assumptions (e.g., independence assumptions, noiseless settings, model well-specification, asymptotic approximations only holding locally). The authors should reflect on how these assumptions might be violated in practice and what the implications would be.
        \item The authors should reflect on the scope of the claims made, e.g., if the approach was only tested on a few datasets or with a few runs. In general, empirical results often depend on implicit assumptions, which should be articulated.
        \item The authors should reflect on the factors that influence the performance of the approach. For example, a facial recognition algorithm may perform poorly when image resolution is low or images are taken in low lighting. Or a speech-to-text system might not be used reliably to provide closed captions for online lectures because it fails to handle technical jargon.
        \item The authors should discuss the computational efficiency of the proposed algorithms and how they scale with dataset size.
        \item If applicable, the authors should discuss possible limitations of their approach to address problems of privacy and fairness.
        \item While the authors might fear that complete honesty about limitations might be used by reviewers as grounds for rejection, a worse outcome might be that reviewers discover limitations that aren't acknowledged in the paper. The authors should use their best judgment and recognize that individual actions in favor of transparency play an important role in developing norms that preserve the integrity of the community. Reviewers will be specifically instructed to not penalize honesty concerning limitations.
    \end{itemize}

\item {\bf Theory assumptions and proofs}
    \item[] Question: For each theoretical result, does the paper provide the full set of assumptions and a complete (and correct) proof?
    \item[] Answer: \answerNA{} 
    \item[] Justification: We show no theoretical results.
    \item[] Guidelines:
    \begin{itemize}
        \item The answer \answerNA{} means that the paper does not include theoretical results. 
        \item All the theorems, formulas, and proofs in the paper should be numbered and cross-referenced.
        \item All assumptions should be clearly stated or referenced in the statement of any theorems.
        \item The proofs can either appear in the main paper or the supplemental material, but if they appear in the supplemental material, the authors are encouraged to provide a short proof sketch to provide intuition. 
        \item Inversely, any informal proof provided in the core of the paper should be complemented by formal proofs provided in appendix or supplemental material.
        \item Theorems and Lemmas that the proof relies upon should be properly referenced. 
    \end{itemize}

    \item {\bf Experimental result reproducibility}
    \item[] Question: Does the paper fully disclose all the information needed to reproduce the main experimental results of the paper to the extent that it affects the main claims and/or conclusions of the paper (regardless of whether the code and data are provided or not)?
    \item[] Answer: \answerYes{} 
    \item[] Justification: We have provided detailed descriptions that ensure the information needed to recreate our datasets, model training, and evaluation results is included. We have also linked to code for reproducing our main experimental results.
    \item[] Guidelines:
    \begin{itemize}
        \item The answer \answerNA{} means that the paper does not include experiments.
        \item If the paper includes experiments, a \answerNo{} answer to this question will not be perceived well by the reviewers: Making the paper reproducible is important, regardless of whether the code and data are provided or not.
        \item If the contribution is a dataset and\slash or model, the authors should describe the steps taken to make their results reproducible or verifiable. 
        \item Depending on the contribution, reproducibility can be accomplished in various ways. For example, if the contribution is a novel architecture, describing the architecture fully might suffice, or if the contribution is a specific model and empirical evaluation, it may be necessary to either make it possible for others to replicate the model with the same dataset, or provide access to the model. In general. releasing code and data is often one good way to accomplish this, but reproducibility can also be provided via detailed instructions for how to replicate the results, access to a hosted model (e.g., in the case of a large language model), releasing of a model checkpoint, or other means that are appropriate to the research performed.
        \item While NeurIPS does not require releasing code, the conference does require all submissions to provide some reasonable avenue for reproducibility, which may depend on the nature of the contribution. For example
        \begin{enumerate}
            \item If the contribution is primarily a new algorithm, the paper should make it clear how to reproduce that algorithm.
            \item If the contribution is primarily a new model architecture, the paper should describe the architecture clearly and fully.
            \item If the contribution is a new model (e.g., a large language model), then there should either be a way to access this model for reproducing the results or a way to reproduce the model (e.g., with an open-source dataset or instructions for how to construct the dataset).
            \item We recognize that reproducibility may be tricky in some cases, in which case authors are welcome to describe the particular way they provide for reproducibility. In the case of closed-source models, it may be that access to the model is limited in some way (e.g., to registered users), but it should be possible for other researchers to have some path to reproducing or verifying the results.
        \end{enumerate}
    \end{itemize}

\item {\bf Open access to data and code}
    \item[] Question: Does the paper provide open access to the data and code, with sufficient instructions to faithfully reproduce the main experimental results, as described in supplemental material?
    \item[] Answer: \answerYes{} 
    \item[] Justification: We have made available code for performing dataset creation, training our models, and performing our main evaluations. This, along with the provided instructions in this work are sufficient to reproduce our main experimental results. All datasets were built with data sourced from public databases and can be recreated with the details provided in the text and the code provided.
    \item[] Guidelines:
    \begin{itemize}
        \item The answer \answerNA{} means that paper does not include experiments requiring code.
        \item Please see the NeurIPS code and data submission guidelines (\url{https://neurips.cc/public/guides/CodeSubmissionPolicy}) for more details.
        \item While we encourage the release of code and data, we understand that this might not be possible, so \answerNo{} is an acceptable answer. Papers cannot be rejected simply for not including code, unless this is central to the contribution (e.g., for a new open-source benchmark).
        \item The instructions should contain the exact command and environment needed to run to reproduce the results. See the NeurIPS code and data submission guidelines (\url{https://neurips.cc/public/guides/CodeSubmissionPolicy}) for more details.
        \item The authors should provide instructions on data access and preparation, including how to access the raw data, preprocessed data, intermediate data, and generated data, etc.
        \item The authors should provide scripts to reproduce all experimental results for the new proposed method and baselines. If only a subset of experiments are reproducible, they should state which ones are omitted from the script and why.
        \item At submission time, to preserve anonymity, the authors should release anonymized versions (if applicable).
        \item Providing as much information as possible in supplemental material (appended to the paper) is recommended, but including URLs to data and code is permitted.
    \end{itemize}

\item {\bf Experimental setting/details}
    \item[] Question: Does the paper specify all the training and test details (e.g., data splits, hyperparameters, how they were chosen, type of optimizer) necessary to understand the results?
    \item[] Answer: \answerYes{} 
    \item[] Justification: Details are provided at a high level in the main text, with complete details in the appendices.
    \item[] Guidelines:
    \begin{itemize}
        \item The answer \answerNA{} means that the paper does not include experiments.
        \item The experimental setting should be presented in the core of the paper to a level of detail that is necessary to appreciate the results and make sense of them.
        \item The full details can be provided either with the code, in appendix, or as supplemental material.
    \end{itemize}

\item {\bf Experiment statistical significance}
    \item[] Question: Does the paper report error bars suitably and correctly defined or other appropriate information about the statistical significance of the experiments?
    \item[] Answer: \answerYes{} 
    \item[] Justification: Uncertainty in experimental results is included in all tables and figures where appropriate along with descriptions of the method for its calculation and its meaning. The details of the significance test in section 4.3 are provided.
    \item[] Guidelines:
    \begin{itemize}
        \item The answer \answerNA{} means that the paper does not include experiments.
        \item The authors should answer \answerYes{} if the results are accompanied by error bars, confidence intervals, or statistical significance tests, at least for the experiments that support the main claims of the paper.
        \item The factors of variability that the error bars are capturing should be clearly stated (for example, train/test split, initialization, random drawing of some parameter, or overall run with given experimental conditions).
        \item The method for calculating the error bars should be explained (closed form formula, call to a library function, bootstrap, etc.)
        \item The assumptions made should be given (e.g., Normally distributed errors).
        \item It should be clear whether the error bar is the standard deviation or the standard error of the mean.
        \item It is OK to report 1-sigma error bars, but one should state it. The authors should preferably report a 2-sigma error bar than state that they have a 96\% CI, if the hypothesis of Normality of errors is not verified.
        \item For asymmetric distributions, the authors should be careful not to show in tables or figures symmetric error bars that would yield results that are out of range (e.g., negative error rates).
        \item If error bars are reported in tables or plots, the authors should explain in the text how they were calculated and reference the corresponding figures or tables in the text.
    \end{itemize}

\item {\bf Experiments compute resources}
    \item[] Question: For each experiment, does the paper provide sufficient information on the computer resources (type of compute workers, memory, time of execution) needed to reproduce the experiments?
    \item[] Answer: \answerYes{} 
    \item[] Justification: We report FLOPs required for model training, GPUs used, and training time.
    \item[] Guidelines:
    \begin{itemize}
        \item The answer \answerNA{} means that the paper does not include experiments.
        \item The paper should indicate the type of compute workers CPU or GPU, internal cluster, or cloud provider, including relevant memory and storage.
        \item The paper should provide the amount of compute required for each of the individual experimental runs as well as estimate the total compute. 
        \item The paper should disclose whether the full research project required more compute than the experiments reported in the paper (e.g., preliminary or failed experiments that didn't make it into the paper). 
    \end{itemize}
    
\item {\bf Code of ethics}
    \item[] Question: Does the research conducted in the paper conform, in every respect, with the NeurIPS Code of Ethics \url{https://neurips.cc/public/EthicsGuidelines}?
    \item[] Answer: \answerYes{} 
    \item[] Justification: We follow the NeurIPS Code of Ethics in all respects.
    \item[] Guidelines:
    \begin{itemize}
        \item The answer \answerNA{} means that the authors have not reviewed the NeurIPS Code of Ethics.
        \item If the authors answer \answerNo, they should explain the special circumstances that require a deviation from the Code of Ethics.
        \item The authors should make sure to preserve anonymity (e.g., if there is a special consideration due to laws or regulations in their jurisdiction).
    \end{itemize}

\item {\bf Broader impacts}
    \item[] Question: Does the paper discuss both potential positive societal impacts and negative societal impacts of the work performed?
    \item[] Answer: \answerNo{} 
    \item[] Justification: Our works applications are to the better understanding of bacterial genomes, and while this does have societal impacts they would be distantly related to what we demonstrate and difficult to conceive. Our model is not generative, and thus the impacts of synthetic biology are not discussed.
    \item[] Guidelines:
    \begin{itemize}
        \item The answer \answerNA{} means that there is no societal impact of the work performed.
        \item If the authors answer \answerNA{} or \answerNo, they should explain why their work has no societal impact or why the paper does not address societal impact.
        \item Examples of negative societal impacts include potential malicious or unintended uses (e.g., disinformation, generating fake profiles, surveillance), fairness considerations (e.g., deployment of technologies that could make decisions that unfairly impact specific groups), privacy considerations, and security considerations.
        \item The conference expects that many papers will be foundational research and not tied to particular applications, let alone deployments. However, if there is a direct path to any negative applications, the authors should point it out. For example, it is legitimate to point out that an improvement in the quality of generative models could be used to generate Deepfakes for disinformation. On the other hand, it is not needed to point out that a generic algorithm for optimizing neural networks could enable people to train models that generate Deepfakes faster.
        \item The authors should consider possible harms that could arise when the technology is being used as intended and functioning correctly, harms that could arise when the technology is being used as intended but gives incorrect results, and harms following from (intentional or unintentional) misuse of the technology.
        \item If there are negative societal impacts, the authors could also discuss possible mitigation strategies (e.g., gated release of models, providing defenses in addition to attacks, mechanisms for monitoring misuse, mechanisms to monitor how a system learns from feedback over time, improving the efficiency and accessibility of ML).
    \end{itemize}
    
\item {\bf Safeguards}
    \item[] Question: Does the paper describe safeguards that have been put in place for responsible release of data or models that have a high risk for misuse (e.g., pre-trained language models, image generators, or scraped datasets)?
    \item[] Answer: \answerNA{} 
    \item[] Justification: Our data and models do not have a high risk for misuse.
    \item[] Guidelines:
    \begin{itemize}
        \item The answer \answerNA{} means that the paper poses no such risks.
        \item Released models that have a high risk for misuse or dual-use should be released with necessary safeguards to allow for controlled use of the model, for example by requiring that users adhere to usage guidelines or restrictions to access the model or implementing safety filters. 
        \item Datasets that have been scraped from the Internet could pose safety risks. The authors should describe how they avoided releasing unsafe images.
        \item We recognize that providing effective safeguards is challenging, and many papers do not require this, but we encourage authors to take this into account and make a best faith effort.
    \end{itemize}

\item {\bf Licenses for existing assets}
    \item[] Question: Are the creators or original owners of assets (e.g., code, data, models), used in the paper, properly credited and are the license and terms of use explicitly mentioned and properly respected?
    \item[] Answer: \answerYes{} 
    \item[] Justification: All original assets used are properly credited. Licenses and links to each pre-trained model used are provided as well as links to datasets along with their terms of use.
    \item[] Guidelines:
    \begin{itemize}
        \item The answer \answerNA{} means that the paper does not use existing assets.
        \item The authors should cite the original paper that produced the code package or dataset.
        \item The authors should state which version of the asset is used and, if possible, include a URL.
        \item The name of the license (e.g., CC-BY 4.0) should be included for each asset.
        \item For scraped data from a particular source (e.g., website), the copyright and terms of service of that source should be provided.
        \item If assets are released, the license, copyright information, and terms of use in the package should be provided. For popular datasets, \url{paperswithcode.com/datasets} has curated licenses for some datasets. Their licensing guide can help determine the license of a dataset.
        \item For existing datasets that are re-packaged, both the original license and the license of the derived asset (if it has changed) should be provided.
        \item If this information is not available online, the authors are encouraged to reach out to the asset's creators.
    \end{itemize}

\item {\bf New assets}
    \item[] Question: Are new assets introduced in the paper well documented and is the documentation provided alongside the assets?
    \item[] Answer: \answerYes{} 
    \item[] Justification: We have linked to documented code for training C3P models and performing our evaluations. We have also provided a link for downloading our trained C3P models, with documentation. Our datasets and models are well-documented in the text.
    \item[] Guidelines:
    \begin{itemize}
        \item The answer \answerNA{} means that the paper does not release new assets.
        \item Researchers should communicate the details of the dataset\slash code\slash model as part of their submissions via structured templates. This includes details about training, license, limitations, etc. 
        \item The paper should discuss whether and how consent was obtained from people whose asset is used.
        \item At submission time, remember to anonymize your assets (if applicable). You can either create an anonymized URL or include an anonymized zip file.
    \end{itemize}

\item {\bf Crowdsourcing and research with human subjects}
    \item[] Question: For crowdsourcing experiments and research with human subjects, does the paper include the full text of instructions given to participants and screenshots, if applicable, as well as details about compensation (if any)? 
    \item[] Answer: \answerNA{} 
    \item[] Justification: Our work does not involve crowdsourcing or research with human subjects.
    \item[] Guidelines:
    \begin{itemize}
        \item The answer \answerNA{} means that the paper does not involve crowdsourcing nor research with human subjects.
        \item Including this information in the supplemental material is fine, but if the main contribution of the paper involves human subjects, then as much detail as possible should be included in the main paper. 
        \item According to the NeurIPS Code of Ethics, workers involved in data collection, curation, or other labor should be paid at least the minimum wage in the country of the data collector. 
    \end{itemize}

\item {\bf Institutional review board (IRB) approvals or equivalent for research with human subjects}
    \item[] Question: Does the paper describe potential risks incurred by study participants, whether such risks were disclosed to the subjects, and whether Institutional Review Board (IRB) approvals (or an equivalent approval/review based on the requirements of your country or institution) were obtained?
    \item[] Answer: \answerNA{} 
    \item[] Justification: Our work does not involve crowdsourcing or research with human subjects.
    \item[] Guidelines:
    \begin{itemize}
        \item The answer \answerNA{} means that the paper does not involve crowdsourcing nor research with human subjects.
        \item Depending on the country in which research is conducted, IRB approval (or equivalent) may be required for any human subjects research. If you obtained IRB approval, you should clearly state this in the paper. 
        \item We recognize that the procedures for this may vary significantly between institutions and locations, and we expect authors to adhere to the NeurIPS Code of Ethics and the guidelines for their institution. 
        \item For initial submissions, do not include any information that would break anonymity (if applicable), such as the institution conducting the review.
    \end{itemize}

\item {\bf Declaration of LLM usage}
    \item[] Question: Does the paper describe the usage of LLMs if it is an important, original, or non-standard component of the core methods in this research? Note that if the LLM is used only for writing, editing, or formatting purposes and does \emph{not} impact the core methodology, scientific rigor, or originality of the research, declaration is not required.
    \item[] Answer: \answerNA{} 
    \item[] Justification: The core method development in this research does not involve LLMs as any important, original, or non-standard components.
    \item[] Guidelines:
    \begin{itemize}
        \item The answer \answerNA{} means that the core method development in this research does not involve LLMs as any important, original, or non-standard components.
        \item Please refer to our LLM policy in the NeurIPS handbook for what should or should not be described.
    \end{itemize}

\end{enumerate}

\end{document}